\documentclass[11pt,a4paper]{article}
\usepackage[utf8]{inputenc}
\usepackage[T1]{fontenc}
\usepackage{lmodern}
\usepackage[english]{babel}
\usepackage[margin=25mm]{geometry}
\usepackage{amsmath,amssymb,bm}
\usepackage{graphicx}
\usepackage{placeins}
\usepackage[font=small,labelfont=bf]{caption}
\usepackage{microtype}
\usepackage{xurl}
\usepackage{needspace}
\usepackage[hidelinks]{hyperref}
\title{TetraCultures of Neurons, Astrocytes, Oligodendrocytes, and Microglia:
A Murine \emph{In Vitro} Model to Dissect Region-Specific CNS Immune Dynamics}
\author{\parbox{\dimexpr\textwidth-12pt\relax}{\centering\small
Lea Gabele-Kutsche\textsuperscript{1,2}, Yaning Zhao\textsuperscript{3},
Tabea Spengler\textsuperscript{1}, Kay Michael Gottlieb\textsuperscript{1},
Christian Deppe\textsuperscript{3}, Martin Korte\textsuperscript{1,2,*},
Kristin Michaelsen-Preusse\textsuperscript{1,*}\\[1ex]
\footnotesize
\textsuperscript{1}Department of Cellular Neurobiology, Institute for Cell- and Neurobiology,
TU Braunschweig, 38106 Braunschweig, Germany\\[0.4ex]
\textsuperscript{2}Neuroinflammation and Neurodegeneration Group,
Helmholtz Centre for Infection Research, 38124 Braunschweig, Germany\\[0.4ex]
\textsuperscript{3}Institute for Communications Technology,
TU Braunschweig, 38106 Braunschweig, Germany\\[0.8ex]
\textsuperscript{*}Corresponding authors:
\href{mailto:k.michaelsen@tu-braunschweig.de}{k.michaelsen@tu-braunschweig.de},
\href{mailto:m.korte@tu-braunschweig.de}{m.korte@tu-braunschweig.de}
}}
\date{}
\begin{document}
\maketitle
\begin{abstract}

Neuroinflammation and glial scar formation in the central nervous system
(CNS) emerge from coordinated signaling between neurons, astrocytes,
oligodendrocytes and microglia. Conventional embryonic
neuron-astrocyte-oligodendrocyte (NAO) TriCultures lack microglia and
cannot model innate immune responses. Here, we introduce postnatal
neuron-astrocyte-oligodendrocyte-microglia (NAOM) TetraCultures, in
which resident microglia are retained from the original brain tissue and
co-develop with other CNS cell populations from the onset of culture.
Hippocampal and cortical cultures were maintained for 14 days in vitro
and challenged with LPS, poly I or scratch injury.

Only NAOM TetraCultures mounted robust, region- and stimulus-specific
inflammatory responses. LPS induced marked TNF-\ensuremath{\alpha} and IL-1\ensuremath{\beta} secretion and
increased IBA1 in hippocampal microglia, whereas poly I elicited IFN-\ensuremath{\beta}
release without increasing IBA1. CD68 regulation differed between
hippocampus and cortex, and microglial morphology showed pronounced
amoeboid transitions after LPS. Astrocytes in NAOM TetraCultures
expanded their territorial coverage after LPS, an effect absent in NAO
TriCultures. Following scratch injury, IL-1\ensuremath{\beta} increased, microglia
infiltrated lesion zones and astrocytes formed lesion borders.

Thus, NAOM TetraCultures recapitulate key features of CNS immunity,
glial scar formation and region-specific glial responses while providing
a physiologically relevant microglia-containing model.

\end{abstract}

\hypertarget{introduction}{%
\section{Introduction}\label{introduction}}

The mammalian central nervous system (CNS) is a highly complex milieu in
which neurons, ependymal, endothelia and glial cells - astrocytes,
oligodendrocytes and microglia - interact continuously to maintain
homeostasis, support synaptic function, respond to injury and
orchestrate repair. Over the past decades, \emph{in vitro} culture
platforms have been indispensable for dissecting individual cell-type
biology and intercellular communication. Embryonic neuron--
astrocyte--oligodendrocyte co-cultures (here termed NAO TriCultures) are
widely used to study neurodevelopmental processes, myelination,
synaptogenesis and basic glial physiology (Kaech and Banker 2006).
However, these conventional systems inherently lack microglia, the
resident immune cells of the CNS, which colonize the brain during late
gestation and early postnatal stages and are essential for innate immune
surveillance, synaptic pruning and response to both sterile and
pathogen-associated insults (Colonna and Butovsky 2017; Tejera and
Heneka 2016).

Microglia emerge from primitive myeloid progenitors in the yolk sac,
entering the CNS around embryonic day 9.5 in rodents and expanding
through local proliferation during early postnatal life. Their long-term
survival and functional maintenance \emph{in vivo} depends critically on
colony- stimulating factor 1 receptor (CSF1R) signaling, principally
through the ligand IL-34 in forebrain regions (Wang et al. 2012). In the
absence of IL-34, microglial precursors fail to colonize properly, and
mature microglia undergo apoptosis. Similarly, \emph{in vitro} cultures
lacking IL-34 supplementation exhibit rapid microglial attrition,
precluding reliable study of microglia-glia interactions.

Beyond their role as immune sentinels, microglia exhibit remarkable
regional heterogeneity. Early immunohistochemical surveys documented
higher microglial density and IBA1 immunoreactivity in cortex compared
to hippocampus (Lawson et al. 1990). More recent single-cell
transcriptomic analyses reveal that hippocampal and cortical microglia
differ in baseline gene expression and in their responsiveness to
inflammatory stimuli: hippocampal microglia demonstrate a lower
activation threshold and more robust pro-inflammatory transcriptional
shifts after lipopolysaccharide (LPS) challenge, whereas cortical
microglia maintain elevated basal activation but exhibit distinct
downstream signaling profiles (Masuda et al. 2019). Such regional
differences likely reflect variations in local neuronal activity,
extracellular matrix composition and glia--neuron cross-talk,
underscoring the need for brain-region--specific \emph{in vitro} models.

Astrocytes, the most abundant glial cell type, perform critical
functions in neurotransmitter recycling, blood-brain barrier
maintenance, metabolic support of neurons and modulation of synaptic
activity. In response to injury or infection, astrocytes undergo
reactive astrogliosis - characterized by GFAP upregulation, process
hypertrophy and altered cytokine secretion - which can be both
neuroprotective and detrimental (Schiweck et al. 2021). Importantly,
microglia- derived mediators such as IL-1\ensuremath{\beta} and TNF-\ensuremath{\alpha} amplify astrocytic
hypertrophy and scar formation, whereas the absence of microglia
attenuates these responses (Goshi et al. 2020). Conventional tri- and
co-cultures lacking microglia thus offer an incomplete picture of
astrocyte reactivity and preclude investigation of microglia-dependent
mechanisms of neuroinflammation.

Mechanical insults to the CNS - modeled \emph{in vitro} by scratch
assays - elicit sterile inflammatory responses characterized by
damage-associated molecular pattern (DAMP) release, inflammasome
activation and IL-1\ensuremath{\beta} secretion without necessarily engaging Toll-like
receptor pathways that drive TNF-\ensuremath{\alpha} or IL-6 secretion (Allan and
Rothwell, 2001). \emph{In vivo} and monoculture studies show that IL-1\ensuremath{\beta}
acts as an early alarmin, promoting glial migration and scar formation,
while microglia rapidly adopt amoeboid morphologies to clear debris
(Allan and Rothwell 2001; Lopez-Castejon and Brough 2011). Yet the
interplay among microglia, astrocytes and oligodendrocytes during scar
formation remains understudied \emph{in vitro}, particularly in a
region- specific context.

While (Goshi et al. 2020) pioneered rat neuron-astrocyte-microglia
tricultures to demonstrate microglial necessity for LPS-induced
astrocyte hypertrophy and cytokine release, their model did not address
viral mimetics, region-specific differences or interactions with
oligodendrocytes. Moreover, most platforms use mixed or undefined brain
tissue, obscuring regional heterogeneity. There remains a clear need for
a pair of matched, region-specific culture systems that can isolate
microglia-independent and microglia-dependent processes, support
long-term viability of all four principal CNS cell types, and
recapitulate both pathogen-associated and sterile injury responses.

In addition to experimental characterization, neuroinflammatory
responses can be viewed as multidimensional phenotype patterns composed
of cytokine secretion, activation-marker expression, and morphological
remodeling (Paolicelli et al. 2022; Masuda et al. 2020). Because these
features often change simultaneously in a stimulus-dependent manner,
interpreting the overall response using individual measurements alone
can be challenging (Deczkowska et al. 2018). Computational
representations that organize complex phenotypes into a limited number
of dominant response patterns have become increasingly useful for
analyzing high-dimensional biological systems (Ching et al. 2018;
Luecken and Theis 2019) and may therefore provide a useful framework for
comparing distinct neuroinflammatory perturbations and identifying
common principles of microglial activation. To fill this gap, we
compared two complementary murine culture platforms. First, NAO
TriCultures (neurons, astrocytes, oligodendrocytes) are derived from
embryonic hippocampus or cortex and maintained in defined
Neurobasal-based medium. Second, NAOM TetraCultures incorporate
microglia by harvesting postnatal (P1) hippocampus or cortex and
supplementing the culture medium with IL-34 (50 ng/mL) alongside other
trophic factors. Both platforms are maintained for 14 days \emph{in
vitro}, enabling direct comparison of region- and stimulus-specific
responses.

Using these paired systems, we first confirm maintenance of
physiologically relevant cell ratios and brain-region identity under
basal conditions. We then challenge cultures with bacterial (LPS), viral
(poly I:C) and mechanical (scratch) insults and quantify
pro-inflammatory cytokine release, microglial morphology and activation
markers (IBA1, CD68), astrocyte reactivity (coverage, GFAP) and glial
migration dynamics. By contrasting NAO TriCultures and NAOM
TetraCultures, we dissect microglia-dependent amplification and
modulation of neuroimmune responses and glial scar formation. Finally,
we introduce a computational framework to integrate multidimensional
neuroimmune measurements and describe stimulus-specific microglial
response programs across distinct inflammatory paradigms.

Our paired culture platforms provide a versatile, high-throughput tool
for mechanistic studies of CNS immunity, injury response and repair in a
region-specific context. They open new avenues for screening
anti-inflammatory or pro-repair therapeutics, investigating gene- or
cell-type- specific contributions, and ultimately bridging the gap
between reductionist \emph{in vitro} assays and \emph{in vivo}
complexity. By capturing both microglia-independent and
microglia-dependent phenomena, NAO TriCultures and NAOM TetraCultures
establish a comprehensive \emph{in vitro} framework for elucidating
multicellular mechanisms underlying neuroinflammation and CNS repair.

\hypertarget{results}{%
\section{Results}\label{results}}

\hypertarget{baseline-cellular-composition-in-hippocampal-and-cortical-cultures}{%
\subsection{Baseline Cellular Composition in Hippocampal and Cortical Cultures}\label{baseline-cellular-composition-in-hippocampal-and-cortical-cultures}}

Before probing functional responses, we confirmed that our paired
culture platforms faithfully incorporate intended cell types at
physiologically relevant ratios. To assess baseline composition, we
compared NAOM TetraCultures (postnatal, four-cell) and NAO TriCultures
(embryonic, threecell) from hippocampus or cortex at 14 DIV. In both
regions, neurons comprised the majority of cells (hippocampus: 68\% in
NAOM vs. 58\% in NAO; cortex: 65\% vs. 53\%), with astrocytes accounting
for \ensuremath{\sim}18--20\% in hippocampal and \ensuremath{\sim}20--27\% in cortical preparations.
Microglia represented \ensuremath{\sim}12\% of hippocampal and \ensuremath{\sim}8\% of cortical cells in
NAOM TetraCultures, whereas NAO TriCultures (by design) lacked microglia
and contained oligodendrocytes at 17\% (hippocampus) and 27\% (cortex)
(Fig. 1B, D). Total cell densities were comparable across culture types
(1 400--1 600 cells mm\ensuremath{^-}\ensuremath{^2}), with neuronal densities at \ensuremath{\sim}900--1 100 cells
mm\ensuremath{^-}\ensuremath{^2} and astrocytes at \ensuremath{\sim}300 cells mm\ensuremath{^-}\ensuremath{^2}. Microglia in NAOM TetraCultures
and oligodendrocytes in NAO TriCultures were present at moderate
densities (150--250 cells mm\ensuremath{^-}\ensuremath{^2}), demonstrating that our postnatalderived
medium supports microglial survival without altering overall cell
numbers relative to the conventional NAO TriCultures.

\begin{figure}[!htbp]
\centering
\includegraphics[width=\linewidth,height=0.68\textheight,keepaspectratio]{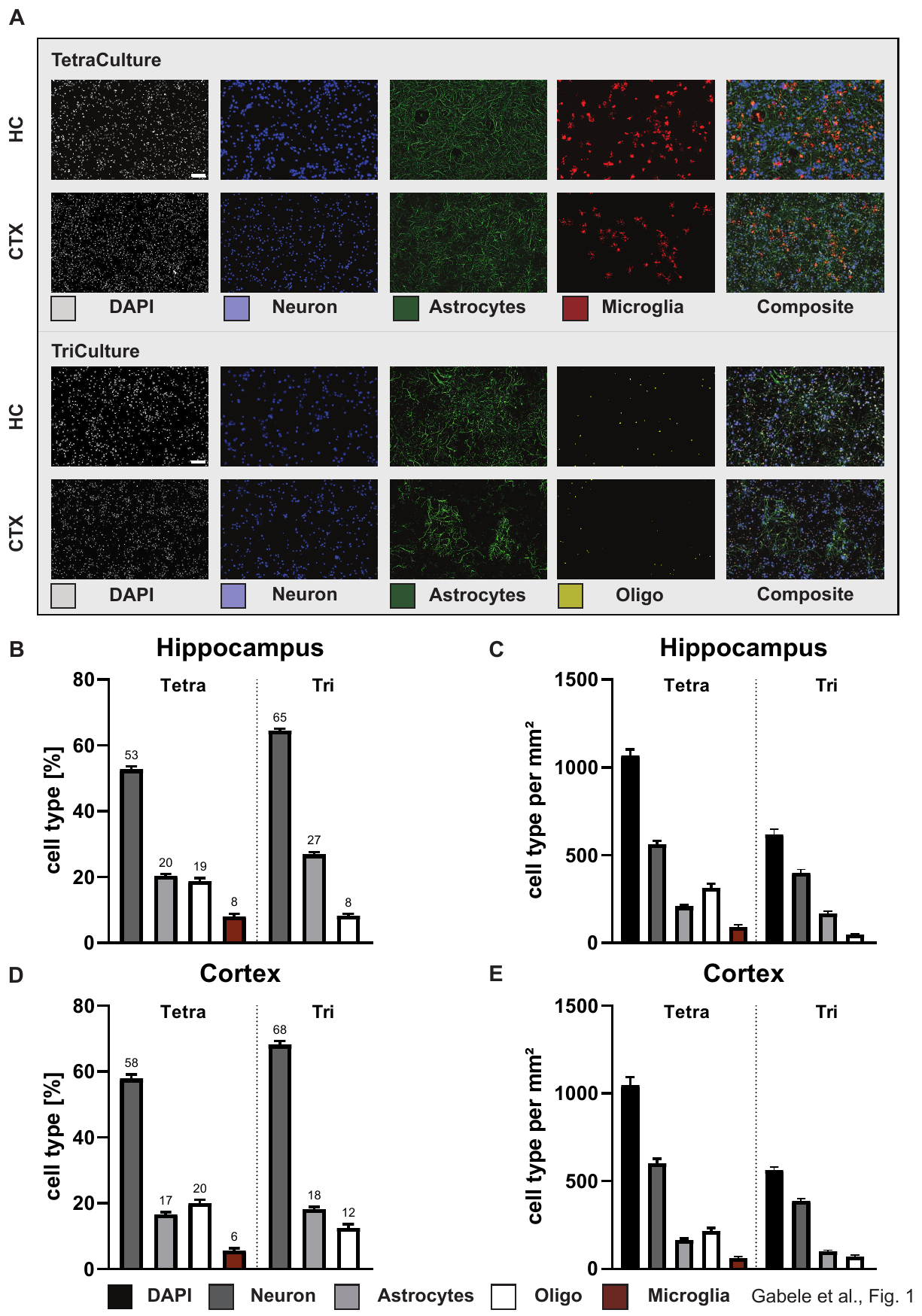}
\caption{\textbf{Composition of hippocampal and cortical NAOM TetraCultures and NAO TriCultures at DIV14.} (A) Representative confocal micrographs of DIV14 hippocampal and
cortical cultures. NAOM TetraCultures were immunolabeled for neurons
(NeuN; blue), astrocytes (GFAP; green) and microglia (IBA1; red). NAO
TriCultures were immunolabeled for neurons (NeuN; blue), astrocytes (GFAP; green) and oligodendrocytes (Olig2; yellow). All panels include
DAPI nuclear counterstain (grey). Scale bar: 50 \ensuremath{\mu}m. (B, D) Relative
proportions of neurons, astrocytes and microglia (B) or oligodendrocytes
(D) in hippocampal (B) and cortical (D) NAOM TetraCultures versus NAO
TriCultures, expressed as percentage of total cells. (C, E) Absolute
cell densities (cells mm\ensuremath{^-}\ensuremath{^2}) for each cell type and total cell number in
hippocampal (C) and cortical (E) cultures. N=3; n=60 (images, 10 per
round).}
\label{fig:figure_01}
\end{figure}
\FloatBarrier

\hypertarget{stimulus-specific-cytokine-secretion-in-naom-tetra--and-nao-tricultures}{%
\subsection{Stimulus-Specific Cytokine Secretion in NAOM Tetra- and NAO TriCultures}\label{stimulus-specific-cytokine-secretion-in-naom-tetra--and-nao-tricultures}}

We measured IL-6, IL-1\ensuremath{\beta}, TNF-\ensuremath{\alpha} and IFN-\ensuremath{\beta} release after exposing DIV 14
cultures to LPS, poly(I:C) or mechanical scratch. NAOM TetraCultures
showed an \ensuremath{\sim}8-fold hippocampal and \ensuremath{\sim}7-fold cortical increase after LPS,
plus a \ensuremath{\sim}4-fold hippocampal rise with poly(I:C); NAO TriCultures lacked
any poly(I:C) response (Fig. 2A). In NAOM TetraCultures, hippocampal
IL-1\ensuremath{\beta} increased \ensuremath{\sim}6-fold (LPS), \ensuremath{\sim}3-fold (poly(I:C)) and \ensuremath{\sim}2-fold
(scratch); cortical NAOM rose \ensuremath{\sim}4-fold after LPS or scratch only. NAO
TriCultures had negligible IL-1\ensuremath{\beta} under all conditions (Fig. 2B). Both
LPS (\ensuremath{\sim}10-fold) and poly(I:C) (\ensuremath{\sim}5-fold) induced strong TNF-\ensuremath{\alpha} secretion in
NAOM TetraCultures across regions; NAO TriCultures showed minimal
release (Fig. 2C). Poly(I:C) triggered IFN-\ensuremath{\beta} in all conditions, but LPS
or scratch only modestly elevated IFN-\ensuremath{\beta} in NAO TriCultures (Fig. 2D).
These data confirm that only NAOM TetraCultures mount robust, region-
and stimulus-specific innate immune responses.

\begin{figure}[!htbp]
\centering
\includegraphics[width=\linewidth,height=0.68\textheight,keepaspectratio]{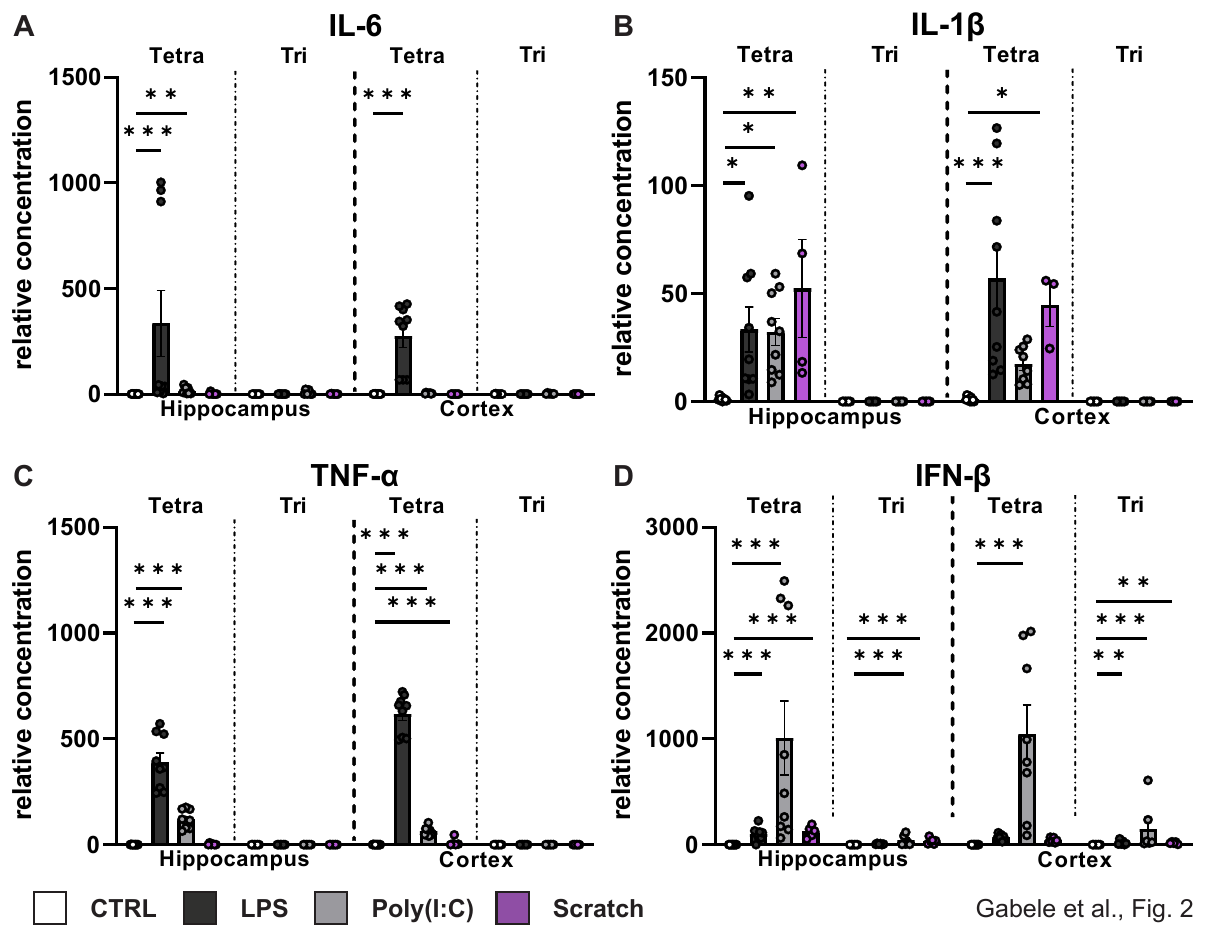}
\caption{\textbf{Pro-inflammatory cytokine release from DIV14 hippocampal and cortical NAOM TetraCultures and NAO TriCultures following immune and mechanical challenges.} NAOM TetraCultures and NAO TriCultures were treated for 24 h with
bacterial lipopolysaccharide (LPS; 1 \ensuremath{\mu}g/mL), viral mimetic poly (I:C)
(50 \ensuremath{\mu}g/mL) or mechanical injury (``Scratch''), then supernatants were
assayed by ELISA. Data are plotted as mean \ensuremath{\pm} SEM (n=3--5 independent
cultures); ``CTRL'' denotes untreated controls. Data were
log\ensuremath{_2}-transformed; significance by two-way ANOVA with Tukey's multiple
comparisons. (A) IL-6 release in hippocampal and cortical NAOM
TetraCultures. (B) IL-1\ensuremath{\beta} release in hippocampal and cortical NAOM
TetraCultures. (C) TNF\ensuremath{\alpha} release in hippocampal and cortical NAOM
TetraCultures. (D) IFN-\ensuremath{\beta} release in both NAOM TetraCultures and NAO
TriCultures from hippocampus and cortex.}
\label{fig:figure_02}
\end{figure}
\FloatBarrier

\hypertarget{morphological-remodeling-of-microglia-following-lps-and-polyic}{%
\subsection{Morphological Remodeling of Microglia Following LPS and poly(I:C)}\label{morphological-remodeling-of-microglia-following-lps-and-polyic}}

We quantified microglial cluster shifts, morphometrics and activation
markers in NAOM TetraCultures. LPS drove hippocampal amoeboid cells from
\ensuremath{\sim}15\% to \ensuremath{\sim}60\%, and cortical from \ensuremath{\sim}15\% to \ensuremath{\sim}50\%; poly(I:C) induced only
minor intermediate increases (Fig. 3B-D). Radar analyses showed LPS
markedly reduced branching, convex hull area and branch length while
increasing circularity and density; poly(I:C) had subtler effects (Fig.
3C-D). LPS upregulated IBA1 by \ensuremath{\sim}1.8-2.2- fold in both phenotypes of
hippocampal and cortical microglia, whereas poly(I:C) reduced IBA1 by
\ensuremath{\sim}30\% (Fig. 3F). LPS left CD68 unchanged or slightly elevated, while
poly(I:C) decreased CD68 by \ensuremath{\sim}25-40\% across both regions (Fig. 3G).
These results demonstrate distinct pro- versus subdued activation states
in response to bacterial versus viral mimetics.

\begin{figure}[!htbp]
\centering
\includegraphics[width=\linewidth,height=0.64\textheight,keepaspectratio]{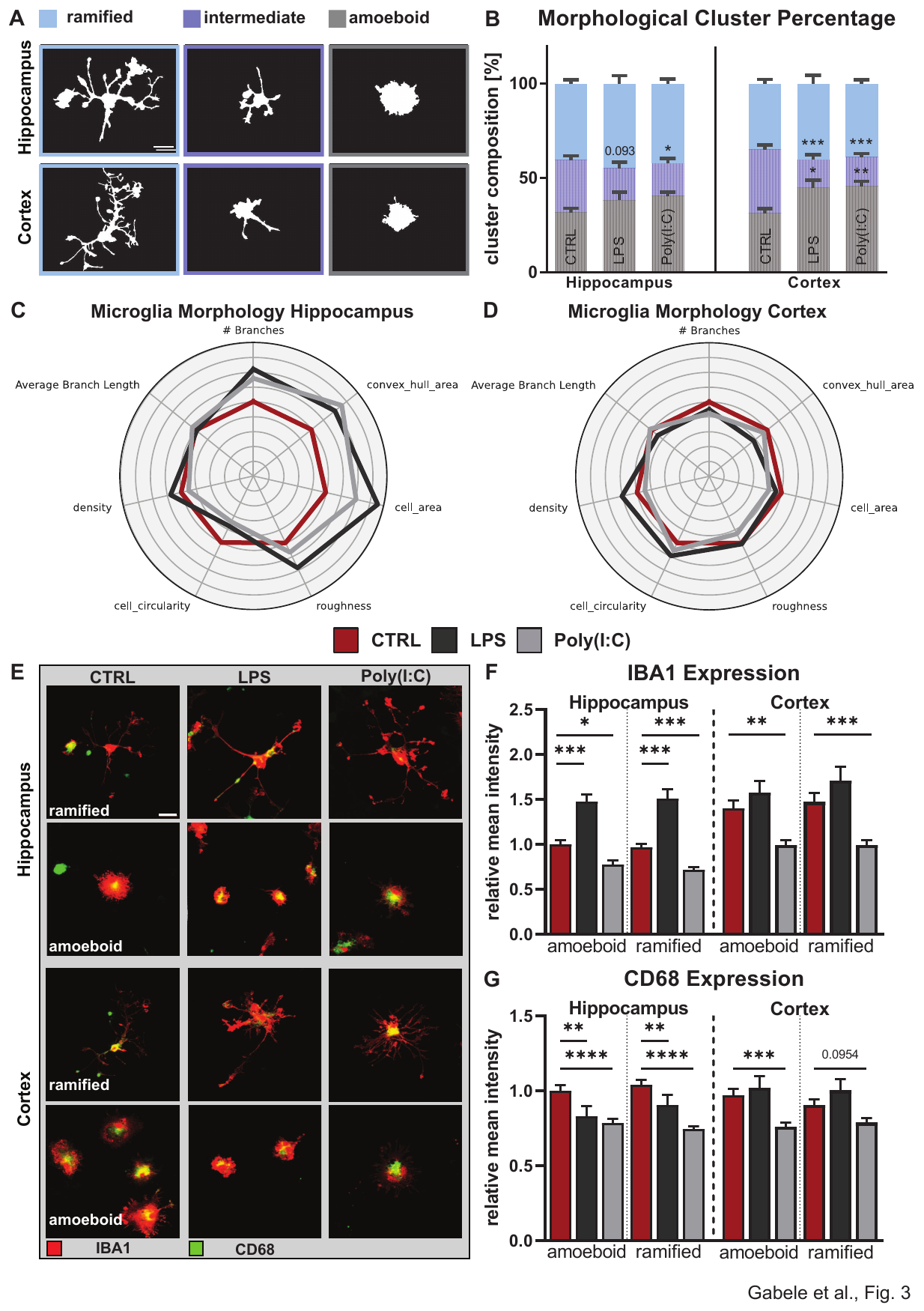}
\caption{\textbf{Microglial morphology and activation marker expression in DIV14 hippocampal and cortical NAOM TetraCultures following bacterial and viral mimetic stimulation.} NAOM TetraCultures were treated for 24 h with LPS (1 \ensuremath{\mu}g/mL), poly(I:C)
(50 \ensuremath{\mu}g/mL) or left untreated (``Ctrl''), then analyzed by
immunofluorescence and image processing. All data are normalized to
hippocampal NAOM TetraCulture control and plotted as mean \ensuremath{\pm} SEM. (A) Representative binarized images of IBA1-stained microglia (scale
bar: 20 \ensuremath{\mu}m). (B) Distribution of microglial morphological clusters -
ramified, intermediate and amoeboid---after stimulation, in hippocampal
and cortical NAOM TetraCultures. (C, D) Radar plots showing relative
changes in seven morphological parameters (branch number, convex hull
area, cell area, roughness, circularity, density and average branch
length) in hippocampal (C) and cortical (D) NAOM TetraCultures. (E)
Confocal images of hippocampal and cortical microglia in amoeboid versus
ramified states, co-stained for IBA1 (red) and CD68 (green) (scale bar:
20 \ensuremath{\mu}m). (F) Quantification of IBA1 mean fluorescence intensity in
amoeboid and ramified microglia. (G) Quantification of CD68 mean
fluorescence intensity in amoeboid and ramified microglia. Cluster
analysis data were subjected to a one-way ANOVA followed by Sidak's
multiple comparisons. IBA1 and CD68 expression data were not normally
distributed and therefore subjected to a Kruskal-Wallis test followed by
a Dunn's multiple comparisons test. N=3, n=30-60 (images, 10 per round
for LPS, 20 per round for PolyIC).}
\label{fig:figure_03}
\end{figure}
\FloatBarrier

\hypertarget{region-dependent-astrocyte-reactivity-to-bacterial-and-viral-mimetics}{%
\subsection{Region-Dependent Astrocyte Reactivity to Bacterial and Viral Mimetics}\label{region-dependent-astrocyte-reactivity-to-bacterial-and-viral-mimetics}}

We assessed astrocyte coverage and GFAP expression in both culture
types. Under basal conditions, NAOM TetraCultures exhibited \ensuremath{\sim}1.4-fold
greater astrocyte area than NAO TriCultures in cortex. LPS further
increased astrocyte coverage in NAOM (+25\% hippocampus, +35\% cortex)
but decreased it in NAO (\ensuremath{\sim}15\% hippocampus, \ensuremath{\sim}20\% cortex). Poly(I:C) had
no effect in NAOM but modestly reduced coverage (\ensuremath{\sim}18\%) in cortical NAO
TriCultures only (Fig. 4B). Baseline GFAP was lower in hippocampal NAOM
(0.8-fold) but higher in cortical NAOM (1.2-fold) versus NAO. LPS
decreased GFAP in both hippocampal (\ensuremath{\sim}30\% NAOM; --25\% NAO) and cortical
NAOM (\ensuremath{\sim}20\%). Poly(I:C) decreased GFAP in cortical NAOM (\ensuremath{\sim}15\%) but
increased it in cortical NAO (+20\%), with no hippocampal impact (Fig.
4C). These findings reveal that microglia presence shapes both astrocyte
expansion and molecular reactivity in a region- and stimulus-specific
manner.

\begin{figure}[!htbp]
\centering
\includegraphics[width=\linewidth,height=0.68\textheight,keepaspectratio]{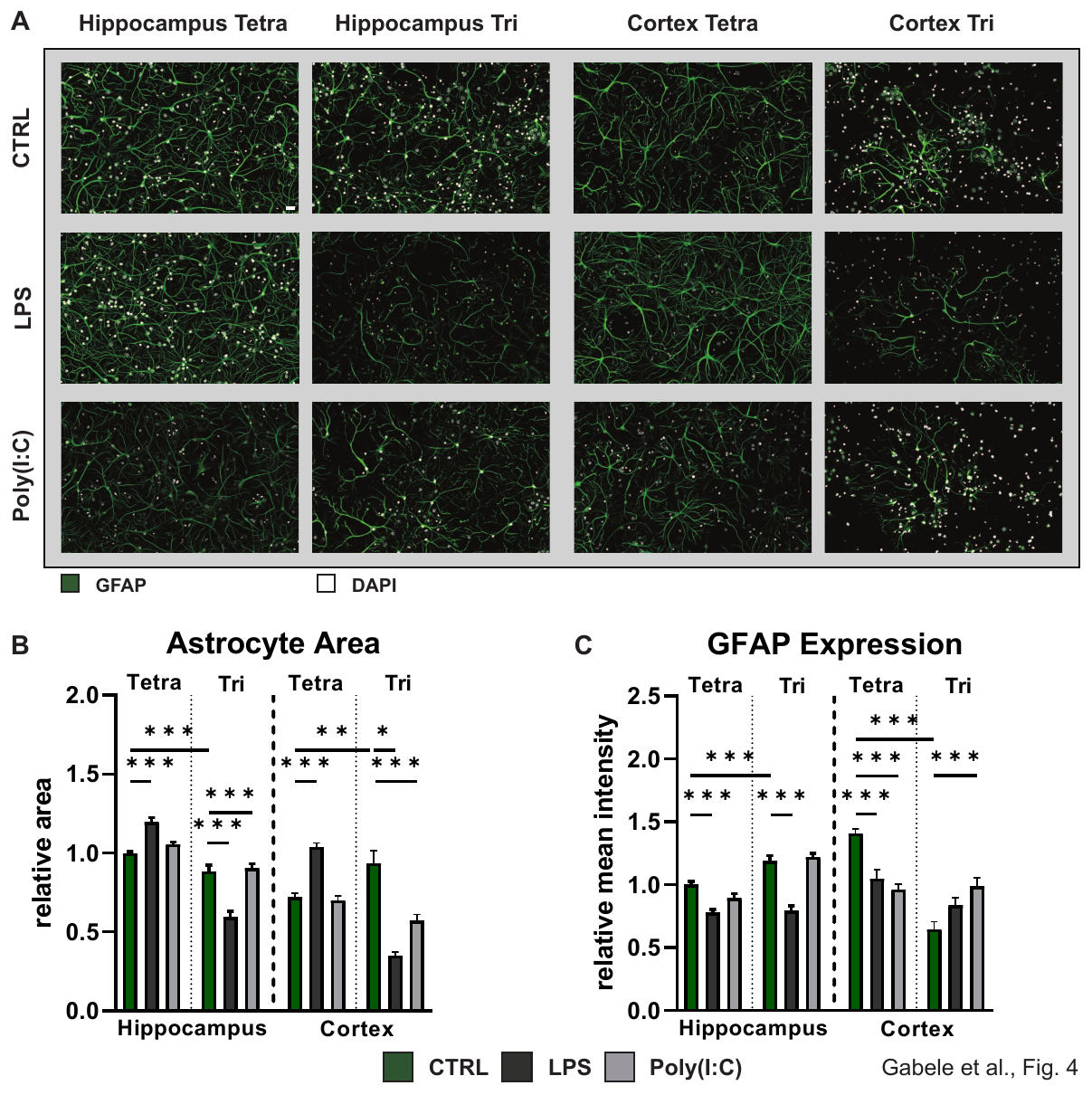}
\caption{\textbf{Region-specific astrocyte reactivity in DIV14 hippocampal and cortical NAOM TetraCultures and NAO TriCultures following bacterial (LPS) or viral (poly (I:C)) challenge.} NAOM TetraCultures and NAO TriCultures were treated for 24 h with LPS (1
\ensuremath{\mu}g/mL), poly(I:C) (50 \ensuremath{\mu}g/mL) or left untreated (Ctrl), then fixed and
analyzed. All data are normalized to the hippocampal NAOM TetraCulture
control and are presented as mean \ensuremath{\pm} SEM. (A) Representative confocal
images of GFAP-stained astrocytes (green) with DAPI nuclear counterstain
(grey) in hippocampal and cortical NAOM TetraCultures and NAO
TriCultures (scale bar: 20 \ensuremath{\mu}m). (B) Quantification of astrocyte-covered area (relative to Ctrl) in
  hippocampal and cortical cultures.
(C)
  Mean GFAP fluorescence intensity (relative to Ctrl) in hippocampal and
  cortical NAOM TetraCultures versus NAO TriCultures. Astrocyte GFAP
  expression data were normally distributed and subjected to a two-way
  ANOVA followed by Tukey's multiple comparisons. Astrocyte area data
  were not normally distributed and therefore subjected to a
  Kruskal-Wallis test followed by a Dunn's multiple comparisons test
  N=3, n=30-60 (images, 10 per round for LPS, 20 per round for PolyIC).
}
\label{fig:figure_04}
\end{figure}
\FloatBarrier

\hypertarget{coordinated-glial-migration-into-lesion-zones-after-mechanical-injury}{%
\subsection{Coordinated Glial Migration into Lesion Zones after Mechanical Injury}\label{coordinated-glial-migration-into-lesion-zones-after-mechanical-injury}}

Scratch assays quantified astrocyte and microglial invasion (Figure 5A).
NAO TriCultures exhibited greater astrocyte coverage in the scratch than
NAOM TetraCultures (hippocampus: 1.8-fold; cortex: 2.2-fold); NAOM
maintained more coverage at borders. Cortical NAO had minimal
infiltration (\textless10\%) (Fig. 5B). GFAP intensity inside the
scratch was 1.5-fold higher than border in hippocampal NAOM, whereas
cortical NAOM border GFAP exceeded NAO (Fig. 5C). Lesion densities were
elevated versus borders in both regions (hippocampus: 3.0 vs. 0.8;
cortex: 2.5 vs. 0.7 cells per \ensuremath{\mu}m\ensuremath{^2}) and in both culture types. Microglial
area within scratches was \ensuremath{\sim}2.5-fold higher than borders (Fig. 5D-E).
These data recapitulate early scar formation, highlighting composition-
and region-dependent migratory dynamics.

\begin{figure}[!htbp]
\centering
\includegraphics[width=\linewidth,height=0.68\textheight,keepaspectratio]{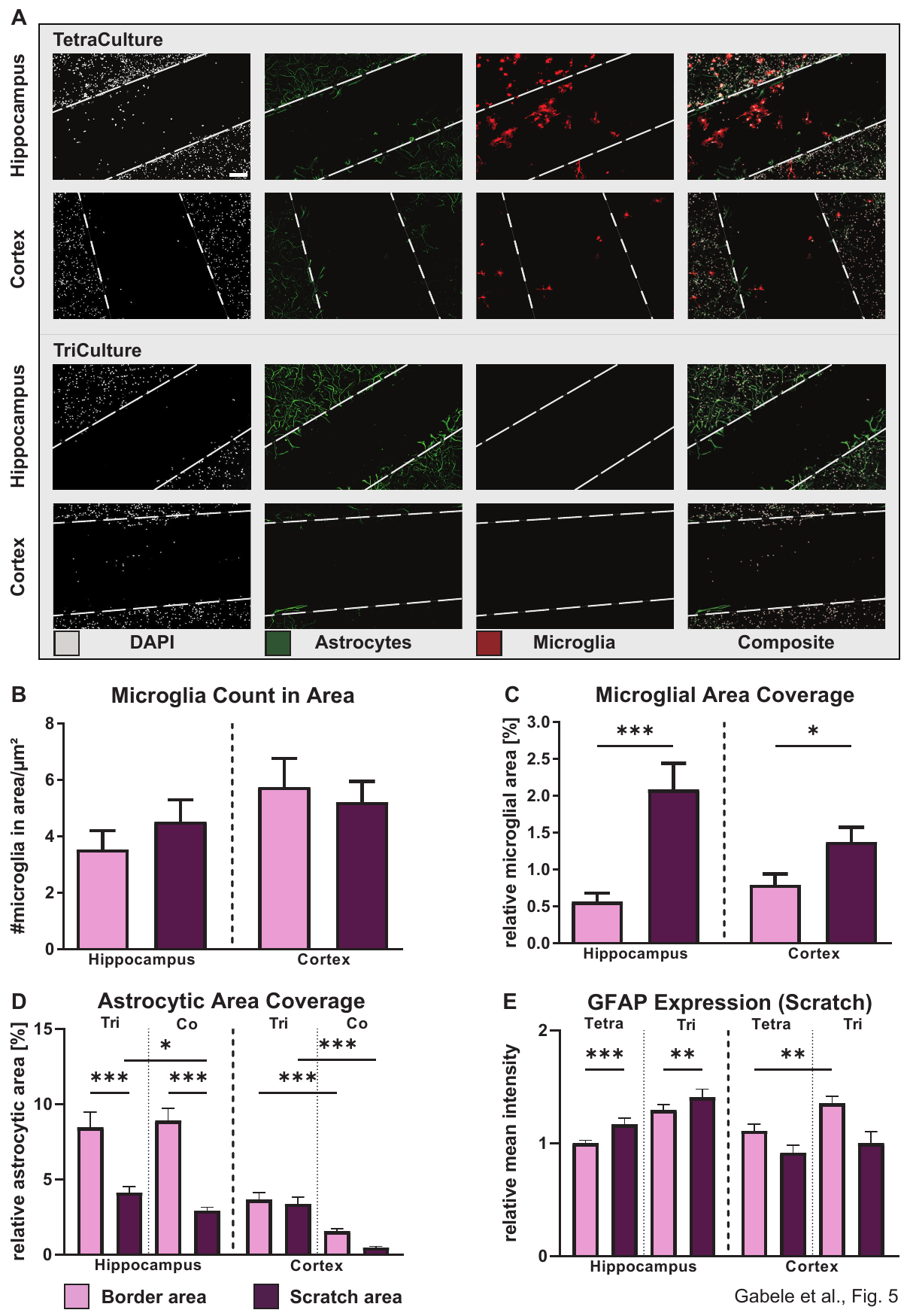}
\caption{\textbf{Reactive glial migration in DIV14 hippocampal and cortical NAOM TetraCultures and NAO TriCultures following mechanical injury.} NAOM TetraCultures and NAO TriCultures were subjected to a standardized
``scratch'' injury and incubated for 24 h before fixation and analysis.
All data are presented as mean \ensuremath{\pm} SEM. (A) Representative low-magnification micrographs showing DAPI (grey),
GFAP-positive astrocytes (green) and IBA1-positive microglia (red)
migrating toward the scratch in hippocampal and cortical NAOM
TetraCultures and NAO TriCultures (scale bar: 100 \ensuremath{\mu}m). (B) Relative
astrocyte-covered area (\%) within the scratch zone versus adjacent
border region. (C) GFAP fluorescence intensity (mean ROI intensity) in
astrocytes located inside the scratch or in the border, normalized to
hippocampal NAOM TetraCulture border control. (D) Microglial density (\#
microglia \ensuremath{\mu}m\ensuremath{^-}\ensuremath{^2}) in scratch and border areas. (E) Relative
microglial-covered area (\%) in scratch versus border regions. An
unpaired two-tailed t-test was used to analyze microglia data, a two-way
ANOVA with Tukey's multiple comparisons test was used for astrocyte
data. N=3, n=30 (images, 10 per round).}
\label{fig:figure_05}
\end{figure}
\FloatBarrier

\hypertarget{local-microglial-transformation-in-response-to-scratch}{%
\subsection{Local Microglial Transformation in Response to Scratch}\label{local-microglial-transformation-in-response-to-scratch}}

We analyzed microglial morphology and markers post-scratch in NAOM
TetraCultures (Fig. 6A). Amoeboid cells in scratches rose from \ensuremath{\sim}20\% to
\ensuremath{\sim}65\%, with border distributions unchanged (Fig. 6B). Hippocampal
microglia showed marked decreases in branching and area and increases in
circularity and density; cortical shifts were similar but milder (Fig.
6C-D). IBA1 intensity increased \ensuremath{\sim}1.5- to 2-fold in both phenotypes and
regions (Fig. 6F). Hippocampal CD68 remained stable, whereas cortical
border microglia upregulated CD68 by \ensuremath{\sim}30\%, with no change inside
scratches (Fig. 6G). These findings demonstrate that mechanical injury
induces a localized, region-specific amoeboid transformation and
activation-marker modulation in microglia.

\begin{figure}[!htbp]
\centering
\includegraphics[width=\linewidth,height=0.64\textheight,keepaspectratio]{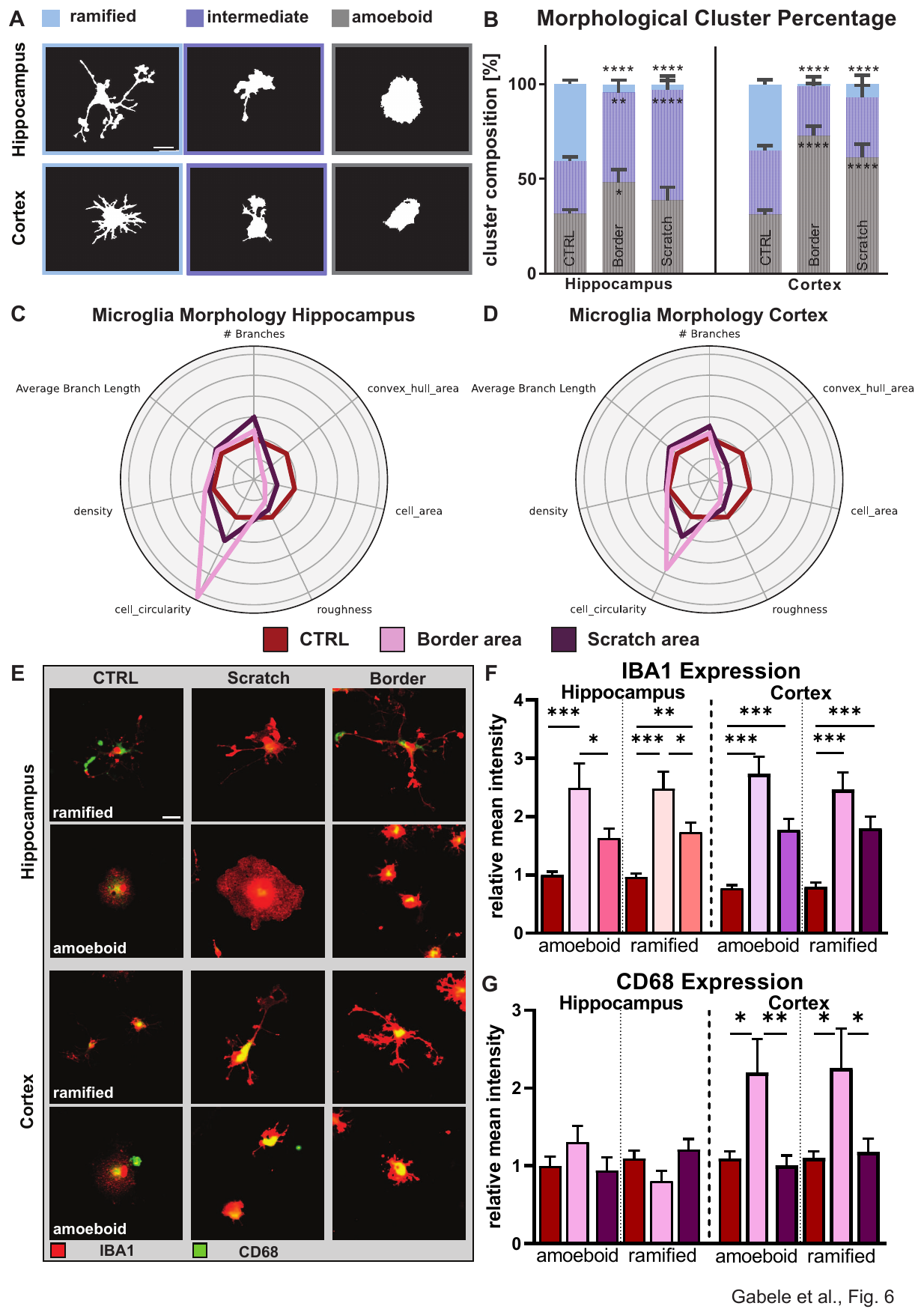}
\caption{\textbf{Microglial morphology and activation marker expression in DIV14 hippocampal and cortical NAOM TetraCultures following mechanical injury (scratch assay).} NAOM TetraCultures were subjected to a standardized scratch wound and
incubated for 24 h. All measures are normalized to the hippocampal NAOM
TetraCulture control and presented as mean \ensuremath{\pm} SEM. (A) Binarized
IBA1-stained microglia illustrating morphology in control (Ctrl), border
and scratch zones (scale bar: 20 \ensuremath{\mu}m). (B) Proportion of microglia in
ramified, intermediate and amoeboid clusters under control versus
scratch conditions, separately quantified in border and scratch regions.
(C, D) Radar plots of seven morphometric parameters---branch number,
convex hull area (CHA), cell area, roughness, circularity, density and
average branch length---in hippocampal (C) and cortical (D) NAOM
TetraCultures. (E) Images of amoeboid versus ramified microglia in
hippocampus and cortex, co-stained for IBA1 (red) and CD68 (green)
(scale bar: 20 \ensuremath{\mu}m). (F) Relative IBA1 fluorescence intensity in amoeboid
and ramified microglia following scratch injury. (G) Relative CD68
fluorescence intensity in amoeboid and ramified microglia after scratch
injury, showing region-specific changes. Kruskal-Wallis test with Dunn's
multiple comparisons test and One-Way-ANOVA with Šídák's multiple
comparisons test, respectively. N=3, n=30 (images, 10 per round)}
\label{fig:figure_06}
\end{figure}
\FloatBarrier

\hypertarget{computational-framework-of-stimulus-specific-microglial-response-programs}{%
\subsection{Computational framework of stimulus-specific microglial response programs}\label{computational-framework-of-stimulus-specific-microglial-response-programs}}

The distinct cytokine and morphological signatures observed following
LPS stimulation, poly(I:C) stimulation, and scratch injury indicated
that microglial activation in NAOM TetraCultures is not adequately
represented by a single activation state. Instead, each perturbation
generated a characteristic combination of cytokine secretion,
activation-marker expression, and morphological remodeling.

To provide a quantitative description of these multidimensional response
patterns, we developed a latent-state framework in which extracellular
perturbations are mapped onto a limited number of microglial response
programs that generate the observed phenotype outputs (Fig. 7A).

Three representative response programs were defined based on the
experimentally observed phenotype structure. The pro-inflammatory
program (P) was associated with elevated TNF-(\ensuremath{\alpha}), IL-6, IL-1( \ensuremath{\beta} ), and
increased IBA1 expression. The interferon-dominant program (V) was
characterized by enhanced IFN-(\ensuremath{\beta}) production, whereas the wound-response
program (W) was associated with injury-induced IL-1(\ensuremath{\beta}) expression,
amoeboid morphology, and scratch response (Fig. 7C).

Stimulus-to-program mapping revealed preferential activation of distinct
response programs by different perturbations (Fig. 7B). LPS was
associated primarily with the pro-inflammatory program, poly(I:C) with
the interferon-dominant program, and scratch injury with the
wound-response program.

\begin{figure}[!htbp]
\centering
\includegraphics[width=\linewidth,height=0.68\textheight,keepaspectratio]{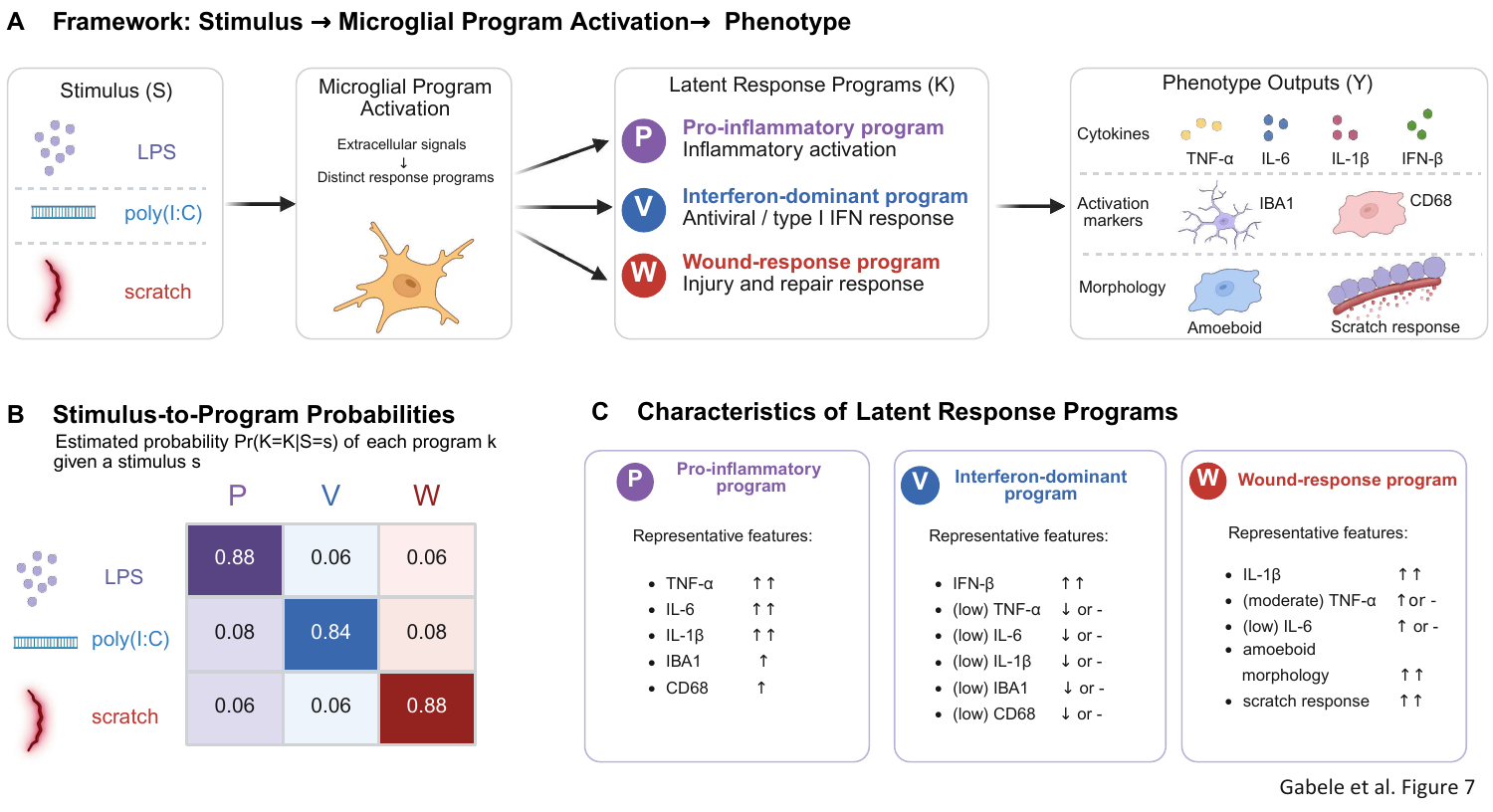}
\caption{\textbf{Framework for stimulus-specific microglial response programs.} The figure summarizes the latent-state framework linking
neuroinflammatory stimuli to microglial response programs and their
characteristic phenotype outputs. (A) Conceptual representation of the
proposed framework. Distinct neuroinflammatory perturbations, including
LPS stimulation, poly(I:C) stimulation, and scratch injury, induce
microglial program activation that is represented by three latent
response programs: pro-inflammatory (P), interferon-dominant (V), and
wound-response (W). Activation of these programs generates
characteristic phenotype outputs, including cytokine production,
activation-marker expression, and morphological responses. (B)
Representative stimulus-to-program activation probabilities. The
conditional probability matrix Pr(\ensuremath{K} = \ensuremath{k},\textbar \ensuremath{S} = \ensuremath{s}) illustrates the
relative activation of each latent response program under different
stimulation conditions. LPS predominantly activates the pro-inflammatory
program (P), poly(I:C) preferentially activates the interferon-dominant
program (V), and scratch injury primarily activates the wound-response
program (W). (C) Characteristics of the latent response programs. The
pro-inflammatory program is characterized by elevated TNF-\ensuremath{\alpha}, IL-6,
IL-1\ensuremath{\beta}, IBA1, and CD68 expression. The interferon-dominant program is
associated primarily with enhanced IFN-\ensuremath{\beta} production accompanied by
comparatively weak inflammatory activation. The wound-response program
is characterized by injury-associated IL-1\ensuremath{\beta} expression, amoeboid
morphology, and the scratch response.}
\label{fig:figure_07}
\end{figure}
\FloatBarrier

\hypertarget{latent-state-model-reconstructs-the-observed-phenotype-structure}{%
\subsection{Latent-state model reconstructs the observed phenotype structure}\label{latent-state-model-reconstructs-the-observed-phenotype-structure}}

To evaluate whether the latent-state framework captures the dominant
structure of neuroimmune responses, reconstructed phenotype signatures
were compared with experimentally observed phenotype patterns derived
from cytokine, activation-marker, and morphology measurements.

Distinct response profiles were observed for each stimulation paradigm
(Fig. 8A). LPS stimulation was associated with strong inflammatory
activation, poly(I:C) produced a prominent interferonrelated phenotype,
and scratch injury generated a wound-associated phenotype characterized
by amoeboid morphology and scratch response.

Model reconstruction closely reproduced the dominant structure of the
experimentally observed phenotype matrix (Fig. 8B). High inflammatory
responses following LPS stimulation, the interferon-associated phenotype
induced by poly(I:C), and the wound-associated phenotype elicited by
scratch injury were all preserved in the reconstructed phenotype space.
Residual analysis revealed only minor deviations between observed and
reconstructed phenotype values (Fig. 8C), with most residuals remaining
within \ensuremath{\pm}0.15 normalized response units.

Together, these observations indicate that the dominant neuroimmune
phenotype structure can be represented using a set of latent response
programs despite substantial dimensionality reduction.

\begin{figure}[!htbp]
\centering
\includegraphics[width=\linewidth,height=0.68\textheight,keepaspectratio]{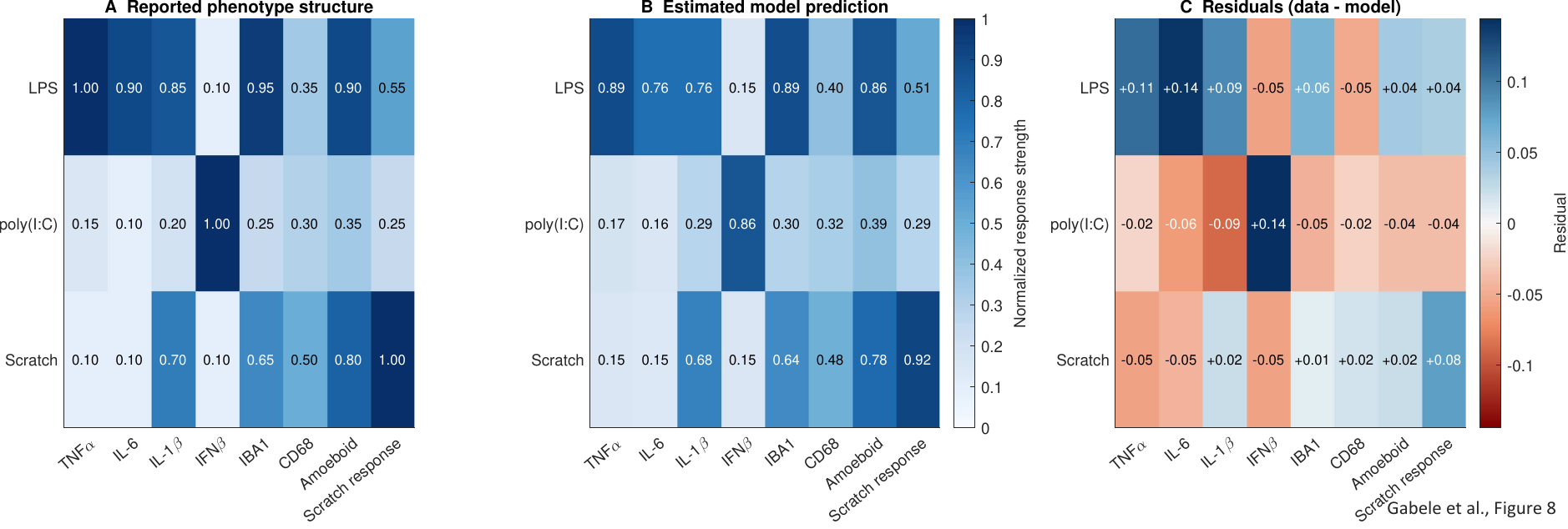}
\caption{\textbf{Reconstruction of experimentally observed phenotype structure using the latent-state framework.} The figure compares the experimentally observed phenotype structure with
its reconstruction by the latent-state framework and shows the
corresponding residuals. (A) Experimentally observed phenotype matrix
based on normalized cytokine, activation-marker, and morphology
measurements following LPS stimulation, poly(I:C) stimulation, and
scratch injury. Phenotype values were normalized to the range {[}0,1{]}
to facilitate comparison across different response modalities. (B)
Phenotype matrix reconstructed by the latent-state framework. Predicted
phenotype values were obtained by combining the characteristic
signatures of the pro-inflammatory (P), interferon-dominant (V), and
wound-response (W) programs according to their stimulus-specific
activation probabilities. (C) Residual matrix showing the difference
between experimentally observed and reconstructed phenotype values
(Observed - Predicted). Positive residuals indicate underestimation by
the model, whereas negative residuals indicate overestimation. Residual
values are displayed in normalized response units.}
\label{fig:figure_08}
\end{figure}
\FloatBarrier

\hypertarget{quantitative-validation-of-the-latent-state-framework}{%
\subsection{Quantitative validation of the latent-state framework}\label{quantitative-validation-of-the-latent-state-framework}}

To further evaluate the fidelity of the latent-state representation,
reconstructed phenotype values were compared directly with
experimentally observed phenotype measurements across all stimulation
paradigms and response variables.

Observed and reconstructed phenotype values exhibited a strong linear
relationship (Fig. 9), with data points clustering closely around the
line of perfect agreement. Quantitative evaluation yielded a coefficient
of determination (\(R^2\) = 0.962), a Pearson correlation
coefficient (\ensuremath{r} = 0.996), and a RMSE of 0.066.

These results demonstrate that the latent-state framework accurately
reconstructs the experimentally observed phenotype space while
preserving the dominant stimulus-specific response patterns.

\begin{figure}[!htbp]
\centering
\includegraphics[width=\linewidth,height=0.68\textheight,keepaspectratio]{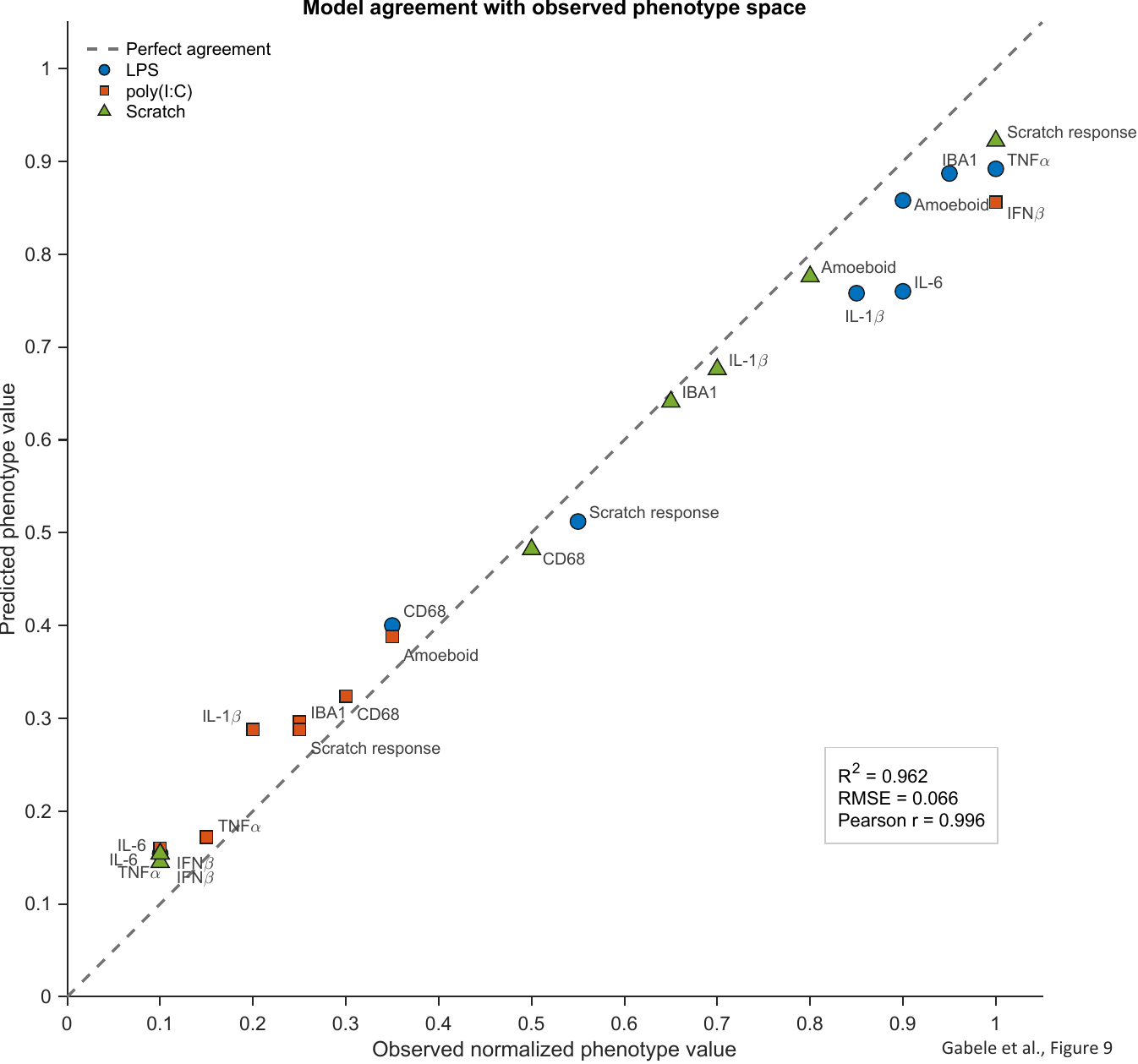}
\caption{\textbf{Quantitative agreement between model predictions and experimentally observed phenotype values.} Observed phenotype values are plotted against the corresponding values
reconstructed by the latent-state framework for all stimulation
paradigms and measured phenotype features. Data points represent
normalized cytokine, activation-marker, and morphology responses
following LPS stimulation, poly(I:C) stimulation, and scratch injury.
The dashed diagonal line indicates perfect agreement between observed
and reconstructed phenotype values. The coefficient of determination (\(R^2\) = 0.962), Pearson
correlation coefficient ((\ensuremath{r} = 0.996)), and root-mean-square error ( RMSE
= 0.066 ) are shown as quantitative measures of reconstruction accuracy.}
\label{fig:figure_09}
\end{figure}
\FloatBarrier

\hypertarget{discussion}{%
\section{Discussion}\label{discussion}}

In this study, we established and validated two complementary murine
culture platforms: the conventional NAO TriCultures, derived from
embryonic (E17.5) hippocampus or cortex and comprising neurons,
astrocytes and oligodendrocytes - an approach widely adopted for
studying neurodevelopmental processes (Kaech and Banker 2006) - and our
newly developed NAOM TetraCultures, which reliably incorporate microglia
by using postnatal (P1) tissue and IL-34 supplementation. By directly
comparing these systems across hippocampal and cortical preparations, we
demonstrate how microglia critically shape neuroimmune and repair
dynamics in a region- and stimulus-specific manner.

Our NAOM TetraCultures successfully incorporated cortical and
hippocampal cell types at physiologically relevant ratios. This was
assessed at 14 days \emph{in vitro} (DIV), a developmental stage at
which the \emph{in vitro} cellular architecture is considered stable and
functionally active. Our results revealed that neurons represented the
majority population in all conditions, with hippocampal cultures
exhibiting 53\% and cortical cultures containing 58\% neuronal content
in NAOM TetraCultures. These values are well-aligned with \emph{in vivo}
estimates, where neurons constitute approximately 60-70\% of cortical
and 50-60\% of hippocampal cells (Erö et al. 2018; Saunders et al. 2018;
Yao et al. 2023), suggesting that our systems preserve regional
distinctions in neuronal density. Astrocytes accounted for approximately
20\% of cells in hippocampal and 17\% in cortical TetraCultures, closely
mirroring their reported \emph{in vivo} abundance of 20-30\% in both
regions (Erö et al. 2018; Yao et al. 2023). Given the crucial roles
astrocytes play in neurotransmitter uptake, metabolic support, and
modulation of synaptic function, their physiological representation
supports the utility of these models for studying neuron-glia
interactions under homeostatic and perturbed conditions. Oligodendrocyte
representation in NAOM TetraCultures (19-20\%) also paralleled values
reported \emph{in vivo}, particularly in cortical regions (Valério-Gomes
et al. 2018). Notably, microglia represented \ensuremath{\sim}6\% of total cells in
cortical and \ensuremath{\sim}8\% in hippocampal NAOM TetraCultures, a range that
corresponds well to physiological densities observed in adult brain
tissue (Erö et al. 2018; Lawson et al. 1990; Yao et al. 2023). These
findings confirm that our defined culture platforms support the
long-term survival of all major CNS cell types and provide a
biologically relevant cellular context for studying region-specific
neuroimmune and glial interactions.

Embryonic NAO TriCultures inherently lack microglia because
yolk-sac-derived precursors invade the CNS during late gestation and
early postnatal stages. As a result, traditional embryonic
neuron-astrocyte-oligodendrocyte cultures are unable to model innate
immune signaling or microglia-astrocyte crosstalk. We overcame this
limitation by harvesting postnatal hippocampal and cortical tissue,
thereby capturing resident microglia alongside the other cell types.
Crucially, we supplemented our culture medium with IL-34 (50 ng/mL), a
high-affinity ligand for the CSF1 receptor that is essential for
microglial survival and proliferation in the forebrain. IL-34's
inclusion replicates \emph{in vivo} trophic support: without it,
microglia rapidly decline \emph{in vitro}, but with IL-34 we maintained
stable microglial densities and functionality for at least 14 days
\emph{in vitro} (Wang et al.

2012). This combination of postnatal origin and IL-34 supplementation is
therefore critical to enabling our NAOM TetraCultures to interrogate
microglia-dependent neuroimmune and repair processes. Additionally, we
supplemented the TetraCulture media with TGF-\ensuremath{\beta} (2ng/mL) and cholesterol
(1.5 mg/mL) to further enhance maintaining physiologically active
microglia (Goshi et al. 2020). TGF-\ensuremath{\beta} has been shown to support the
maintenance of a quiescent microglial phenotype \emph{in vitro} (Abutbul
et al. 2012) and may be crucial for preserving their transcriptional
identity under culture conditions (Butovsky et al. 2014). The addition
of excess cholesterol to the culture medium may further stabilize
microglial gene expression by suppressing apolipoprotein E (ApoE) levels
(Mahley 2016). Although ApoE is essential for lipid transport, its
elevated expression has been negatively associated with the expression
of microglial signature genes (Butovsky et al. 2014), indicating that
modulating lipid metabolism could be a key strategy for maintaining
microglial homeostasis \emph{in vitro}.

Baseline IBA1 expression was consistently higher in cortical NAOM
TetraCultures than in hippocampal, mirroring classic immunohistochemical
findings of greater microglial density and IBA1 immunoreactivity in
rodent cortex compared to hippocampus (Lawson et al. 1990). Upon LPS
challenge, hippocampal microglia exhibited a significant upregulation of
IBA1, whereas cortical microglia did not, suggesting that hippocampal
cells either have a lower activation threshold or more robust TLR4
signalling cascades. Single-cell transcriptomic studies support this
regional sensitivity, showing preferential upregulation of
pro-inflammatory and activationrelated genes in hippocampal microglia
following endotoxin exposure (Masuda et al. 2019). In contrast,
poly(I:C) did not elevate IBA1 in either region, indicating that
TLR3-mediated viral mimetic signaling predominantly engages interferon
pathways without amplifying IBA1 expression - a pattern previously
observed in primary microglial cultures (Colonna and Butovsky 2017; He
et al. 2021).

CD68, a lysosomal marker associated with phagocytic activity, displayed
stimulus- and region-specific regulation in our platform. In hippocampal
cultures, both LPS and poly(I:C) elicited significant downregulation of
CD68, suggesting that sustained inflammatory signaling via either TLR4
or TLR3 may transiently suppress lysosomal biogenesis or turnover in
hippocampal microglia

- a phenomenon noted when microglia shift toward a more secretory, less
phagocytic phenotype (He et al. 2021). By contrast, in cortical cultures
CD68 was only downregulated following poly(I:C), while LPS left CD68
levels unchanged. This divergence may reflect intrinsically higher
baseline CD68 expression in cortical microglia, which resist modulation
by bacterial cues but remain susceptible to interferon-driven repression
via TLR3 pathways. Together, these findings underscore the intrinsic
heterogeneity of microglial basal states and activation dynamics across
brain regions.

Astrocytes in NAOM TetraCultures exhibited markedly different responses
to LPS compared to those in NAO TriCultures. Following LPS treatment,
the astrocyte-covered area in NAOM TetraCultures increased significantly
in both hippocampal and cortical preparations, whereas in NAO
TriCultures, lacking microglia, astrocytic coverage contracted. This
dichotomy indicates that microglia-derived signals, such as IL-1\ensuremath{\beta} and
TNF-\ensuremath{\alpha}, potentiate astrocyte process extension and territorial expansion
in response to bacterial challenge, in line with reports that microglial
cytokines drive astrogliosis and process hypertrophy (Schiweck et al.
2021). In the absence of microglial support, LPS appears to trigger
astrocyte retraction or compromise structural integrity, emphasizing
that astrogliosis is not purely cell-autonomous but critically shaped by
glial crosstalk.

Paradoxically, total GFAP fluorescence intensity declined after LPS
across both regions and culture types, despite the differential
morphological responses. This suggests a cytoskeletal reorganization in
which astrocytes extend thinner, more dynamic processes - redistributing
GFAP over a broader territory but at lower density per fiber - or
potentially undergo proteolytic cleavage of GFAP and transcriptional
downregulation. Such a dissociation between morphological expansion and
marker abundance highlights the complexity of reactive astrocyte states
and underscores the necessity of assessing both structural and molecular
metrics when quantifying astrogliosis.

Our scratch assay data reveal that mechanical injury elicits a selective
cytokine signature and coordinated glial migration. IL-1\ensuremath{\beta} was robustly
secreted following scratch, whereas TNF-\ensuremath{\alpha}, IL-6 and IFN-\ensuremath{\beta} remained at
baseline, consistent with a sterile injury paradigm in which damage-
associated molecular patterns activate the inflammasome pathway in
microglia and astrocytes, driving IL-1\ensuremath{\beta} maturation and release without
broad TLR engagement (Allan and Rothwell 2001). IL-1\ensuremath{\beta} acts as a key
alarmin in sterile CNS injury, promoting astrocyte and microglial
migration into the lesion and coordinating scar formation
(Lopez-Castejon and Brough 2011). The absence of IFN-\ensuremath{\beta} further
underscores that mechanical trauma alone does not mimic viral signals
necessary for type I interferon induction. These findings highlight that
our NAOM TetraCultures can discriminate between distinct inflammatory
triggers - demonstrating a bona fide wound response characterized by
inflammasome activation - thereby providing a platform to dissect the
molecular mechanisms governing sterile injury versus pathogen-driven
inflammation.

Across all assays, hippocampal and cortical cultures exhibited distinct
sensitivities and activation thresholds. Hippocampal microglia, despite
lower basal IBA1, showed greater responsiveness to LPS, whereas cortical
microglia, with higher baseline activation, required different stimuli
for further modulation. Astrocyte reactivity and migration patterns
similarly differed by region, underscoring the importance of
brain-region-tailored \emph{in vitro} systems for accurately modeling
neuroimmune processes.

By pairing the conventional embryonic NAO TriCultures with our
postnatal, IL-34-supplemented NAOM TetraCultures, researchers can
delineate microglia-independent from microglia-dependent mechanisms and
investigate diverse neuroimmune scenarios---including sterile injury
versus pathogen-driven inflammation - within a single experimental
framework. Future refinements might incorporate vascular or blood-brain
barrier components via organ-on-chip technologies, extend culture
durations to study chronic inflammation and remyelination, and adapt
these platforms to human iPSC-derived cells for enhanced translational
relevance.

Beyond the experimental observations, the proposed latent-state
framework provides a compact computational representation of
stimulus-specific neuroimmune responses. By organizing experimentally
observed cytokine, activation-marker, and morphological signatures into
a small number of representative response programs, the framework
captures the dominant structure of the phenotype space while
substantially reducing its dimensionality. The close agreement between
reconstructed and experimentally observed phenotype patterns suggests
that diverse neuroinflammatory perturbations can be represented using a
limited set of canonical microglial response programs.

This representation provides a quantitative perspective on microglial
activation that complements conventional analyses based on individual
cytokines or markers. Rather than interpreting each experimental
measurement independently, the framework summarizes coordinated
phenotype responses at the systems level, thereby facilitating
comparisons across different neuroinflammatory stimuli and experimental
conditions. Such low-dimensional representations may also provide a
foundation for future computational analyses of multicellular
neuroimmune communication and stimulus-specific information processing.
More broadly, this framework illustrates how multidimensional
neuroimmune phenotypes may be represented using latent-state
representations analogous to those commonly employed in communication
and information-processing systems.

Importantly, the proposed framework should be interpreted as a
conceptual abstraction rather than a mechanistic model of intracellular
signaling. The latent response programs do not correspond to discrete
biological cell states but instead provide a compact representation of
the dominant phenotype organization observed across different
stimulation paradigms. Future studies incorporating time-resolved
measurements, larger experimental datasets, and additional stimulation
paradigms may enable further refinement and validation of this
computational framework.

The introduction of murine NAOM TetraCultures represents a critical
advance over traditional embryonic tricultures, providing a versatile,
physiologically relevant platform to probe multicellular mechanisms of
CNS immunity, injury response and repair with faithfully mimicking the
\emph{in vivo} conditions as well as the cellular responses.

\hypertarget{conclusion-and-future-directions}{%
\section{Conclusion and Future Directions}\label{conclusion-and-future-directions}}

In this study, we established a pair of complementary murine in vitro
culture systems that enable the systematic investigation of
multicellular neuroimmune interactions under controlled experimental
conditions. By combining conventional embryonic neuron--astrocyte--
oligodendrocyte (NAO) TriCultures with newly developed postnatal
neuron--astrocyte-- oligodendrocyte--microglia (NAOM) TetraCultures, we
demonstrate that the presence of microglia fundamentally alters
inflammatory signaling, glial communication, and injury responses. The
paired culture platforms capture region-specific differences between
hippocampal and cortical tissue and provide a versatile framework for
studying bacterial, viral, and sterile inflammatory stimuli within a
unified experimental setting.

Our results demonstrate that microglia are essential for generating
physiologically relevant neuroimmune responses in vitro. The NAOM
TetraCultures exhibited robust stimulus-dependent cytokine secretion,
distinct microglial activation states, coordinated astrocyte remodeling,
and localized wound responses that were absent or strongly attenuated in
conventional TriCultures. These findings emphasize that multicellular
communication between neurons and different glial populations is a key
determinant of CNS immune dynamics and should be considered when
modeling neuroinflammatory diseases or tissue repair mechanisms.

Beyond the experimental platform, this work introduces a complementary
computational perspective by representing complex neuroimmune phenotypes
through a small number of latent response programs. Based on the
hippocampal NAOM TetraCulture data, the framework captures
stimulus-specific responses through pro-inflammatory,
interferon-dominant, and wound-response programs. Although this
latent-state framework is intentionally conceptual rather than
mechanistic, it illustrates how multidimensional biological measurements
can be integrated into compact representations that preserve the
dominant characteristics of stimulus-specific immune responses. Such
representations provide an intuitive systems-level description that
complements conventional analyses based on individual cytokines,
activation markers, or morphological features.

The proposed framework naturally opens several directions for future
research. On the experimental side, the culture system can be extended
by incorporating additional cellular components such as endothelial
cells, pericytes, or blood--brain barrier models to further approximate
the in vivo microenvironment. Integration with human iPSC-derived
cultures, organ-on-chip technologies, long-term culture paradigms, and
single-cell transcriptomic or spatial omics approaches could provide
deeper insights into chronic neuroinflammation, neurodegeneration, and
regenerative processes while improving translational relevance.

From a computational perspective, future work may move beyond predefined
response programs toward data-driven latent-state inference using
statistical learning and machine-learning approaches. Time-resolved
measurements could enable dynamic state-space models describing
transitions between neuroimmune response programs, while multimodal
datasets may allow integration of cytokine secretion, gene expression,
morphology, electrophysiology, and imaging into unified computational
representations.

More broadly, the concepts introduced here suggest that neuroimmune
communication may be viewed as an information-processing system in which
extracellular perturbations are encoded into coordinated multicellular
response programs that collectively determine tissue behavior.
Developing rigorous mathematical frameworks to quantify information
flow, robustness, redundancy, and decision-making within multicellular
immune networks represents an exciting long-term research direction.
Such approaches could ultimately bridge systems neuroscience,
immunology, and information theory, providing a quantitative foundation
for understanding how complex cellular communities process environmental
information and coordinate adaptive responses.

Taken together, the combined experimental and computational framework
presented here provides a versatile platform for investigating
multicellular CNS immunity and repair. We anticipate that NAOM
TetraCultures, together with quantitative phenotype representations,
will facilitate mechanistic studies of neuroinflammation, accelerate
therapeutic screening, and stimulate the development of predictive
computational models that connect cellular communication with
tissue-level function.

\subsection*{Data availability statement}

The data supporting the findings of this study are available from the
corresponding author upon reasonable request.

\subsection*{Funding}

This research received no external funding.

\subsection*{Author contribution}

LGK, KMP and MK conceived and designed the research. LGK, KMG, and TS
carried out the experiments and data analysis. YZ and CD analyzed the
data. LGK, KMP, CD, and MK wrote the paper. All authors reviewed the
final manuscript.

\hypertarget{methods}{%
\section{Methods}\label{methods}}

\hypertarget{animals-and-ethics}{%
\subsection{Animals and Ethics}\label{animals-and-ethics}}

All procedures adhered to institutional and national guidelines for
animal care and use and were approved by the local authorities and the
animal welfare officer of the TU Braunschweig. C57BL/6J mice of either
sex were used for all primary cultures. Embryonic cultures were prepared
according to §7 TierSchG (33.19-42502-04-22/00099), postnatal cultures
were prepared according to §4 TierSchG.

\hypertarget{preparation-of-embryonic-nao-tricultures}{%
\subsection{Preparation of embryonic NAO TriCultures}\label{preparation-of-embryonic-nao-tricultures}}

Hippocampi were dissected from E17.5 embryos, dissociated in 0.05\%
trypsin-EDTA (Sigma- Aldrich, T3924) for 25 min at 37 \ensuremath{^\circ}C, then
triturated with a fire-polished Pasteur pipette. Cells were pelleted
(1,500 rpm, 5 min), resuspended in DMEM + 10\% FCS (Gibco, 6195-026;
Capricorn, FCS- 62A), washed three times in Neurobasal (Invitrogen,
21103-049), and plated at 7 \ensuremath{\times} 10\textsuperscript{4} cells/well on
poly-L-lysine-coated coverslips (13 mm, 24-well). Cultures were
maintained in Neurobasal + N2, B27 (Invitrogen, 17504-001) and 0.5 mM
L-glutamine (Invitrogen, 25030-024) at 37 \ensuremath{^\circ}C, 5\% CO\ensuremath{_2}, with 20\% medium
changes weekly (Michaelsen et al. 2010).

\hypertarget{preparation-of-postnatal-hippocampal-and-cortical-naom-tetracultures}{%
\subsection{Preparation of postnatal hippocampal and cortical NAOM TetraCultures}\label{preparation-of-postnatal-hippocampal-and-cortical-naom-tetracultures}}

Hippocampi and neocortices were isolated from P1 pups, dissociated and
plated as above, except at 1 \ensuremath{\times} 10\textsuperscript{5} cells/well. After
initial plating in DMEM + 10\% FCS and washes into Neurobasal + N2, B27
and 0.5 mM L-glutamine, cultures were supplemented with
microglial-supporting factors: ovine wool cholesterol (1.5 \ensuremath{\mu}g/mL;
Avanti, 57-88-5), rmIL-34 (50 ng/mL; R\&D Systems, 5195-ML) and rmTGF-\ensuremath{\beta}1
(2 ng/mL; R\&D Systems, 7666-MB). NAOM TetraCultures were maintained at
37 \ensuremath{^\circ}C, 5\% CO\ensuremath{_2}, with 50\% medium exchanges twice weekly. Due to the
limited shelf life of IL-34 and TGF\ensuremath{\beta}, the TetraCulture medium was made
fresh each week (Goshi et al. 2020).

\hypertarget{immunocytochemistry}{%
\subsection{Immunocytochemistry}\label{immunocytochemistry}}

Cells were fixed in 4\% PFA for 15 min, permeabilized in 0.2\% Triton
X-100 for 30 min, and blocked with 1\% BSA for 5 min. Primary antibodies
were applied for 1 h at 36 \ensuremath{^\circ}C; after PBS washes and a 1\% BSA block,
fluorescent secondaries were incubated for 30 min at 36 \ensuremath{^\circ}C. Nuclei were
counterstained with DAPI (1:1,000) for 5 min. Coverslips were mounted in
Fluorogel (EMS).

Cell Composition (tricultures): IBA1 (rabbit, 1:1,000; Synaptic
Systems), GFAP (chicken, 1:1,000; Invitrogen), NeuN (mouse, 1:500;
Merck) + Cy3, Cy2, Cy5 secondaries (1:500). Cell Composition
(co-cultures): Olig2 (rabbit, 1:1,000; Merck), GFAP, NeuN + secondaries
as above. Activation and Morphology: CD68 (rat, 1:1,000), IBA1
(guinea-pig, 1:2,000), GFAP + appropriate secondaries. Spine Analysis:
Synaptophysin (guinea-pig, 1:1,000), PSD-95 (mouse, 1:1,000) + Cy3/Cy5;
cultures transfected with pm-Citrine-C1.

\hypertarget{chemical-transfection}{%
\subsection{Chemical Transfection}\label{chemical-transfection}}

At DIV12, cultures were transfected with pm-Citrine-C1 (Addgene \#54587)
using Lipofectamine 2000 (Invitrogen, 52887). For each well: 1 \ensuremath{\mu}g DNA +
2 \ensuremath{\mu}L Lipofectamine diluted separately in 50 \ensuremath{\mu}L Neurobasal, combined,
incubated 20 min, then added dropwise for 50 min at 36.5 \ensuremath{^\circ}C. Medium was
restored to 500 \ensuremath{\mu}L per well and cells returned to the incubator for 48
h.

\hypertarget{fluorescence-microscopy}{%
\subsection{Fluorescence Microscopy}\label{fluorescence-microscopy}}

Images were acquired on a Zeiss AxioImager 2 with ApoTome. Microglia
morphology/activation: 20\ensuremath{\times} objective (NA 0.8), z-stacks at 0.35 \ensuremath{\mu}m
increments (0.293 \ensuremath{\mu}m/pixel). Cell composition/scratch assays: 10\ensuremath{\times}
objective (NA 0.3), 0.586 \ensuremath{\mu}m/pixel. Spine analysis: 63\ensuremath{\times} oil objective
(NA 1.4), z-stacks at 0.25 \ensuremath{\mu}m (0.10 \ensuremath{\mu}m/pixel). All exposures and
acquisition settings were kept constant per marker. Analysis was
performed in Fiji (Schindelin et al. 2012).

\hypertarget{image-analysis}{%
\subsection{Image Analysis}\label{image-analysis}}

Cell counts: Maximum-intensity projections; microglia, astrocytes and
oligodendrocytes counted manually; neurons and nuclei via StarDist
(Schmidt et al. 2018) with probability thresholds of 0.65 and 0.60,
respectively. Scratch assays: Scratch ROI drawn manually;
astrocyte/microglial coverage and intensity assessed using binarization
(Huang dark threshold) and custom Fiji macros.

Spine density/morphology: Dendrite length measured with the segmented
line tool; spines counted manually; SpineJ plugin (Levet et al. 2020)
for head/neck metrics; synaptic puncta segmented via unsharp mask,
Gaussian blur subtraction, thresholding, watershed, and particle
analysis.

Microglial morphology clustering: Skeletonization (AnalyzeSkeleton) and
FracLac analysis (Karperien 2007); PCA and K-means clustering
(scikit-learn) to define ramified, intermediate and amoeboid phenotypes.
All image-derived data were processed in Python 3.11 using NumPy and
pandas.

\hypertarget{immune-stimulation}{%
\subsection{Immune stimulation}\label{immune-stimulation}}

Immunstimulation was achived using LPS (bacterial mimetic): 1 \ensuremath{\mu}g/mL
final (SIGMA, L3880) for 48 h or Poly(IC) (viral mimetic): 50 \ensuremath{\mu}g/mL
final (SIGMA, P1530) for 24 h. Untreated or PBS-vehicle controls were
included.

\hypertarget{mechanical-injury-scratch-assay}{%
\subsection{Mechanical Injury (Scratch Assay)}\label{mechanical-injury-scratch-assay}}

At DIV13-15, medium was halved, and a \ensuremath{\sim}200--300 \ensuremath{\mu}m cross scratch was
made with a 20 \ensuremath{\mu}L pipette tip. Cultures were returned to fresh medium
(50\% exchange) and incubated 48 h before analysis (Liang et al. 2007).

\hypertarget{elisa}{%
\subsection{ELISA}\label{elisa}}

Supernatants of Tri- and TetraCultures were collected after poly(I:C)
and LPS stimulation as well as after the scratch assay and stored at
-20\ensuremath{^\circ}C until use. Mouse TNF-\ensuremath{\alpha} DuoSet (DY410), mouse IL- 1\ensuremath{\beta}/IL-1F2 DuoSet
(DY401), mouse IL-6 (DY406), mouse IFN-\ensuremath{\beta} (DY8234-05) ELISA kits (R\&D
Systems) were used to determine the levels of cytokines (TNF-\ensuremath{\alpha}, IL-1\ensuremath{\beta},
IL-6, IFN-\ensuremath{\beta}) in the cell supernatants. In brief, 100 \ensuremath{\mu}l of
cytokine-specific capture antibodies were applied to medium binding
96-well microplates and incubated overnight at room temperature. Plates
were washed three times with 0.5\% v/v Tween20 (Roth, 9127.2) in PBS and
blocked in 300 \ensuremath{\mu}l 1\% BSA (SIGMA, A-7906) in PBS for 1 hour at room
temperature. After being washed again, 100 \ensuremath{\mu}l of the samples and
standards were added and incubated for 2 hours at room temperature.
Plates were then washed three times and incubated with 100 \ensuremath{\mu}l of
detection antibodies for 2 hours at room temperature. Plates were rinsed
and 100 \ensuremath{\mu}l of streptavidin-HRP per well was added and incubated for 20
min at room temperature in the dark. After washing again, the substrate
solution (R\&D Systems, 642736) was added and incubated for 20 minutes,
followed by 50 \ensuremath{\mu}l of the stop solution consisting out of 5\% sulfuric
acid (Applichem, 7664-93-9). The plates were read at 450 nm in an Epoch
BioTek microplate reader using Gen5 software (BioTek, USA). Finally, the
measured optical density of the reaction was compared with the optical
density of the known standard samples to determine the protein
concentration in the samples.

\hypertarget{computational-modeling-and-latent-state-analysis}{%
\subsection{Computational Modeling and Latent-State Analysis}\label{computational-modeling-and-latent-state-analysis}}

To provide a quantitative description of stimulus-specific neuroimmune
responses, a latent-state framework was developed based on the
experimentally observed phenotype patterns in hippocampal NAOM
TetraCultures. Cytokine, activation-marker, and morphology measurements
were first normalized to the range {[}0,1{]} to enable direct comparison
across different response modalities. The phenotype vector consisted of
TNF-\ensuremath{\alpha}, IL-6, IL-1\ensuremath{\beta}, IFN-\ensuremath{\beta}, IBA1, CD68, amoeboid morphology, and the
scratch response. Amoeboid morphology was used as the morphologyrelated
feature to capture the stimulus-induced shift from a ramified toward an
activated microglial phenotype. The scratch response represents the
relative wound-associated response to mechanical injury, including glial
responses within the scratch area and its adjacent border region.

Three representative latent response programs were defined according to
the dominant phenotype patterns observed across the three stimulation
paradigms. The pro-inflammatory program (P), interferon-dominant program
(V), and wound-response program (W) were identified based on their
characteristic cytokine, activation-marker, and morphology signatures.
Rather than representing discrete biological cell states, these latent
programs provide a reduced computational representation of the
experimentally observed phenotype organization.

Each response program was assigned a characteristic phenotype signature
vector, \(\mu_k\) , constructed from the normalized relative
response strengths associated with that program. For each stimulus
condition \ensuremath{s}, the corresponding response-program activation coefficients,
\(\pi_{k\mid s}\) , were estimated from the dominant
experimentally observed phenotype characteristics and normalized such
that \(\sum_k\pi_{k\mid s}=1\). The
reconstructed phenotype vector was calculated as \(\widehat Y(s)=\sum_{k\in\{P,V,W\}}\pi_{k\mid s}\mu_k\), where
\(\mu_k\) denotes the phenotype signature associated with
response program \ensuremath{k}, and \(\pi_{k\mid s}\) represents its
relative activation under stimulus condition \ensuremath{s}.

Model performance was evaluated by comparing reconstructed phenotype
values with experimentally observed phenotype measurements.
Reconstruction accuracy was quantified using residual analysis, the
coefficient of determination \(R^2\), the Pearson
correlation coefficient \ensuremath{r}, and the root-mean-square error (RMSE).

\hypertarget{statistics-and-data-visualization}{%
\subsection{Statistics and Data Visualization}\label{statistics-and-data-visualization}}

All analyses were conducted in GraphPad Prism 7. Data are mean \ensuremath{\pm} SEM.
Two-way ANOVA (culture \ensuremath{\times} treatment) with Tukey post hoc tests assessed
main effects and interactions. One-way ANOVA with Sidak's or
Kruskal--Wallis with Dunn's tests were used for within-group
comparisons. Microglial migration was analyzed by unpaired t-test.
Normality was assessed by Kolmogorov-- Smirnov test; all data that were
not normally distributed were log\ensuremath{_2}-transformed to approximate a normal
distribution and permit valid two-way ANOVA analysis. Significance was
set at \ensuremath{*} \ensuremath{p} \textless{} 0.05, \ensuremath{*}\ensuremath{*} \ensuremath{p} \textless{} 0.01 and \ensuremath{*}\ensuremath{*}\ensuremath{*} \ensuremath{p} \textless{}
0.001. Results of the respective statistical test and means\ensuremath{\pm}SEM can be
found in supplementary Table 1. Figures were prepared in Adobe
Illustrator CS5. To enhance the efficiency and quality of scientific
writing and literature review, several AI-based tools were used. ChatGPT
(OpenAI, model o4-mini) supported the refinement of scientific language,
phrasing, and structure. SciSpace Copilot was employed for contextual
explanations of scientific texts and for verifying conceptual
consistency. Perplexity AI and ChatGPT with advanced research
capabilities (Deep Research) were used to assist in exploring recent
literature, identifying relevant references, and summarizing key
findings. All AI-generated outputs were critically assessed, manually
verified, and edited by the author to ensure scientific accuracy and
authorship integrity.

\par\bigskip

\clearpage
\setcounter{figure}{0}
\renewcommand{\thefigure}{S\arabic{figure}}

\begin{figure}[p]
\section*{Supplementary figures}
\centering
\includegraphics[width=\linewidth,height=0.60\textheight,keepaspectratio]{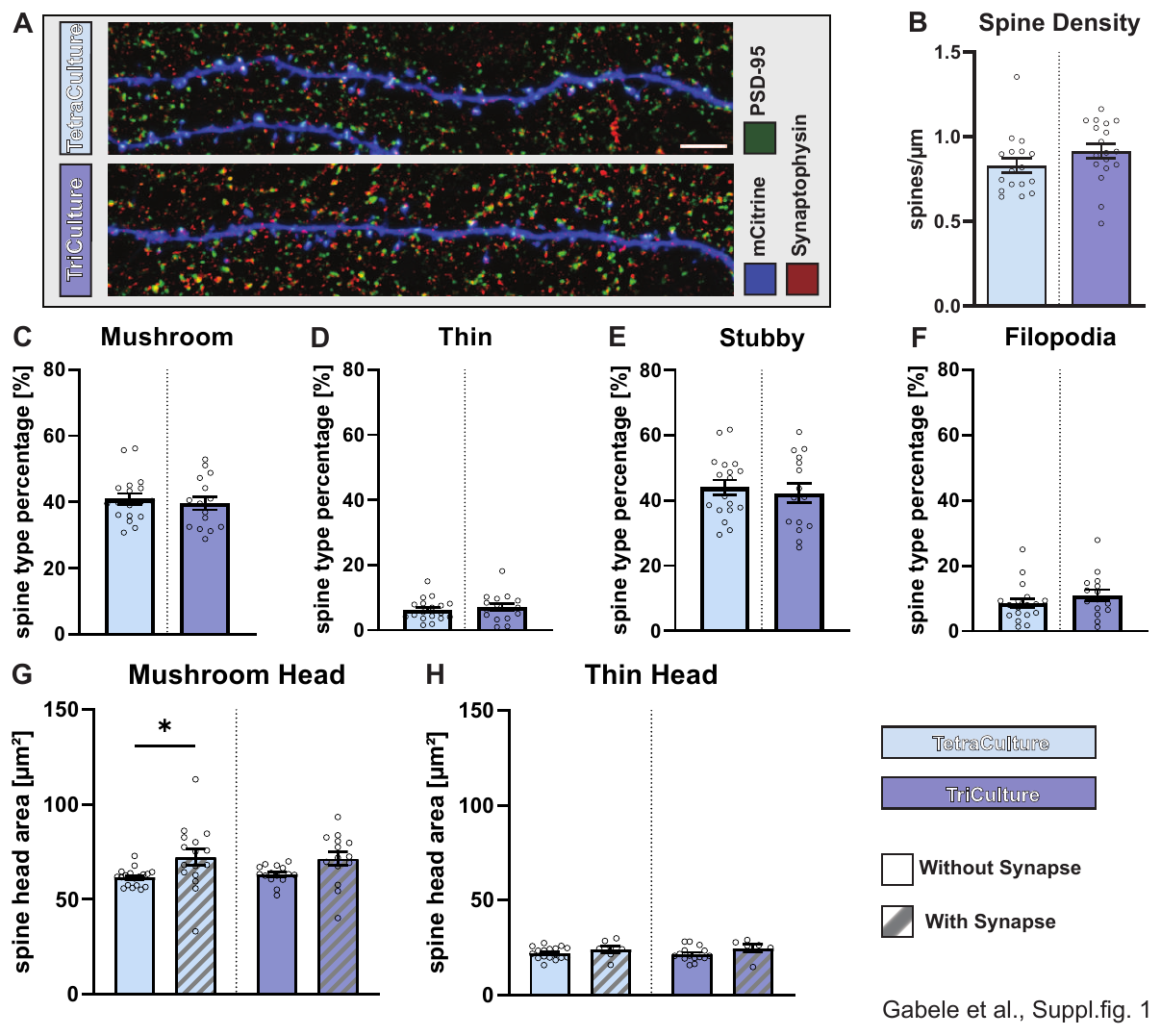}
\caption{\textbf{Dendritic spine metrics of hippocampal neurons in NAO and NAOM cultures.} Quantitative comparisons of dendritic spine metrics in hippocampal
neurons grown under TriCulture versus TetraCulture conditions. All
measures are presented as mean \ensuremath{\pm} SEM. (A) Example images of Tri- and TetraCultures transfected with mCitrine
(blue) and stained for PSD- 95 (green) as well as synaptophysin (red)
(scale bar: 5 \ensuremath{\mu}m). (B) Total spine density of both cultures. Relative
proportion of spine phenotypes in each condition was analyzed: (C)
mushroom spines, (D) thin spines, (E) stubby spines and (F) filopodia-like spines. For
mushroom (G) and thin (H) phenotypes, the spine head area with and without synapse was
analyzed. An unpaired t-test was used to compare Tetra-and TriCulture
spine data. A two-way ANOVA with Tukey's multiple comparisons test was
used for spine head data. N=3, n=18 (images, 6 per round).}
\label{fig:supp_figure_01}
\end{figure}
\clearpage

\begin{figure}[p]
\centering
\includegraphics[width=\linewidth,height=0.68\textheight,keepaspectratio]{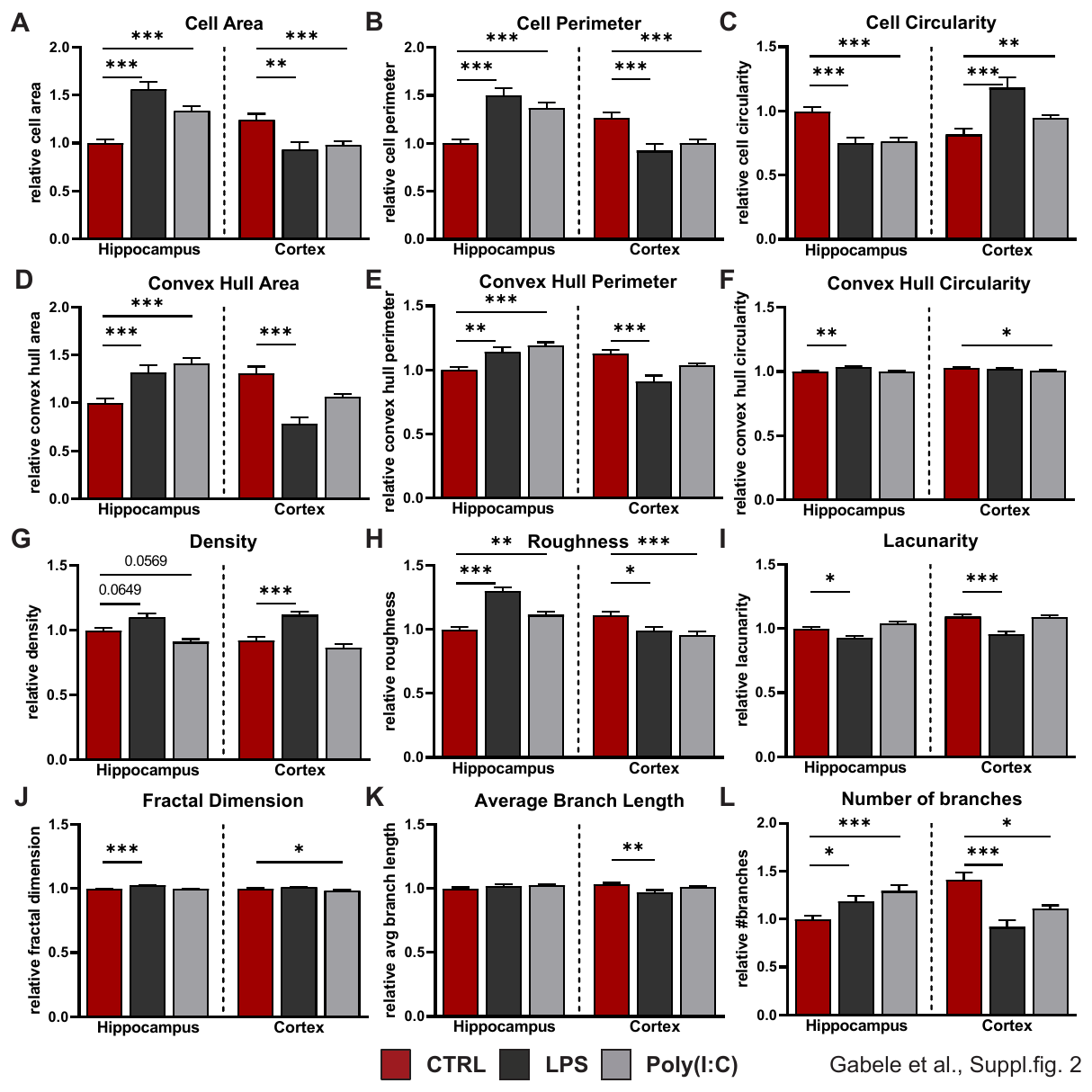}
\caption{\textbf{Morphological characteristics of microglia following neuroinflammatory challenges.} Morphological characteristics of microglia were analyzed using FracLac
(Karperien 2007) and morph data scripts (Campos et al. 2021) in ImageJ.
A self-written python code was subsequently used to perform a K-means
clustering utilizing the Sklearn library to categorize the microglia
into three morphological clusters (as previously described by Gabele et
al., 2024). The characteristics display a morphological change after LPS
stimulation towards an enlarged, more complex, ramified phenotype in
hippocampal microglia, contrarily to a more amoeboid shaped phenotype in
cortical microglia. Following poly(I:C) stimulation, hippocampal
microglial morphological characteristic change towards a more ramified
morphology type, here again contrarily to the cortical microglia, which
morphology changes towards a more roundish-amoeboid phenotype. Cell
complexity is displayed by (H) roughness, (G) density, (J) fractal
dimension and (L) number of branches. The cell size is displayed by (A)
cell area and (D) convex hull area. Generally, the cell shape is
described by (C) cell circularity, (F) convex hull circularity and (K)
average branch length. (B) Cell perimeter as well as (E) convex hull perimeter display both
cell size and complexity. Number of experiments, N=3, n=10 images per
group and cell preparation round. Data are presented as mean \ensuremath{\pm} SEM and
were analyzed by two-way ANOVA followed by post hoc Tukey test.}
\label{fig:supp_figure_02}
\end{figure}
\clearpage

\begin{figure}[p]
\centering
\includegraphics[width=\linewidth,height=0.68\textheight,keepaspectratio]{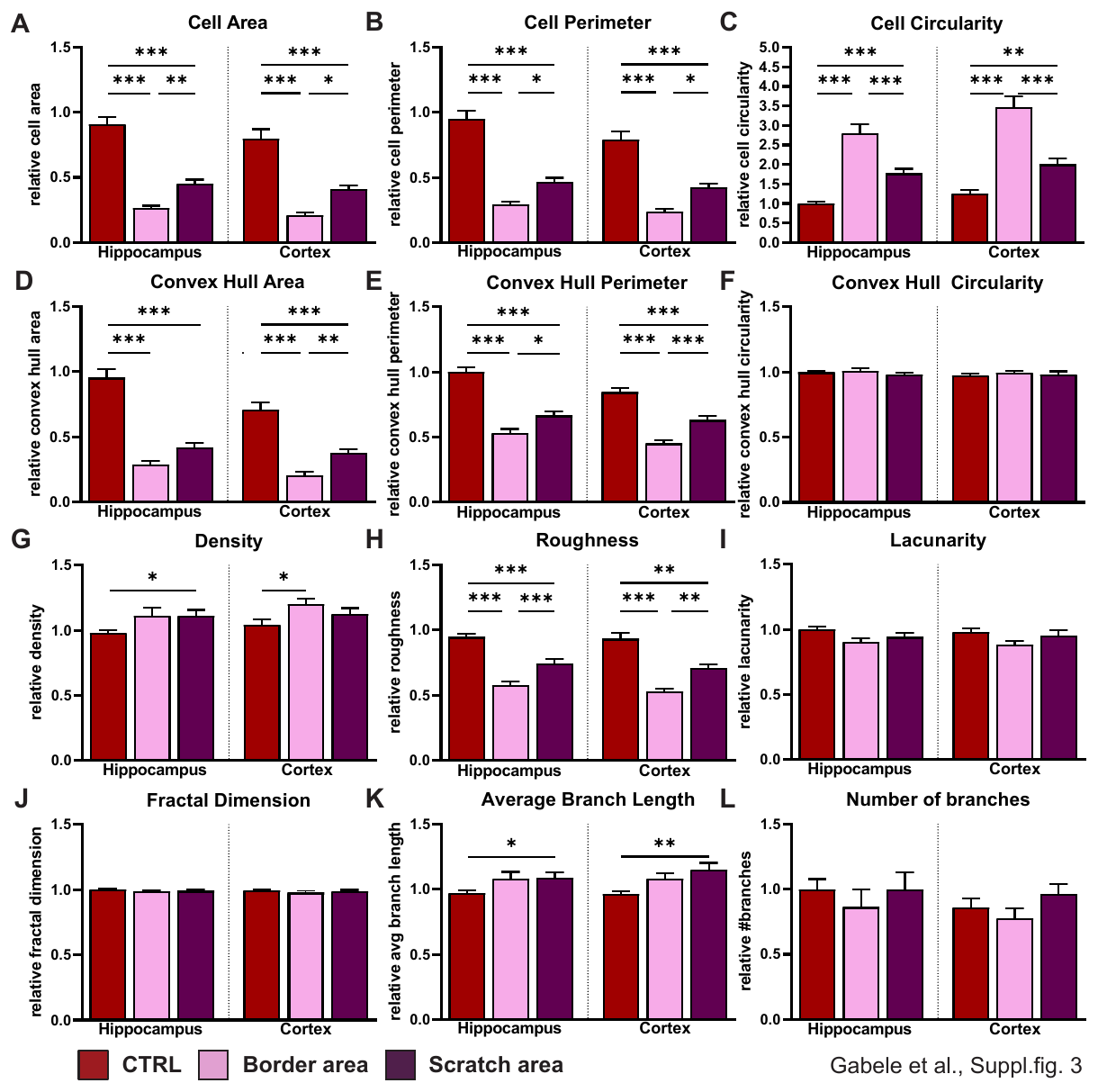}
\caption{\textbf{Morphological characteristics of microglia following mechanical injury.} Morphological characteristics of microglia were analyzed using FracLac
(Karperien 2007) and morph data scripts (Campos et al. 2021) in ImageJ.
A self-written python code was subsequently used to perform a K-means
clustering utilizing the Sklearn library to categorize the microglia
into three morphological clusters (as previously described by Gabele et
al., 2024). The characteristics display a strong morphological shift in
hippocampal as well as cortical microglia after mechanical injury
towards a roundish-amoeboid phenotype indicating microglia activation in
both scratch and border area, whereby the cells seem to be smaller and
more circular directly inside the scratch area. Cell complexity is
displayed by (H) roughness, (G) density, (J) fractal dimension and (L) number of branches. The cell size is displayed by (A) cell area and
(D) convex hull area. Generally, the cell shape is described by (C) cell
circularity, (F) convex hull circularity and (K) average branch length. (B) Cell perimeter as well as (E) convex hull
perimeter displays both cell size and complexity. Number of experiments,
N=3, n=60 (images, 10 per round). Data are presented as mean \ensuremath{\pm} SEM and
were analyzed by two-way ANOVA followed by post hoc Tukey test.}
\label{fig:supp_figure_03}
\end{figure}
\clearpage

\begin{figure}[p]
\centering
\includegraphics[width=\linewidth,height=0.68\textheight,keepaspectratio]{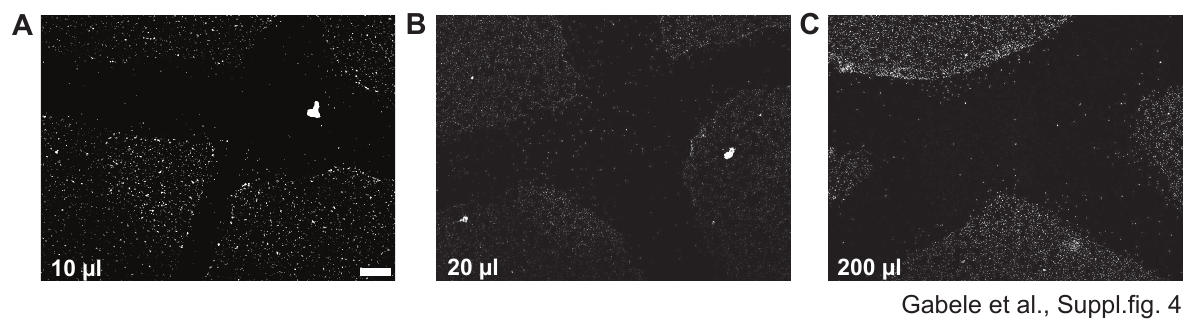}
\caption{\textbf{Overview of pipette tip size used to induce mechanical injury.} For the induction of mechanical injury in terms of a scratch assay, we
used (A) 10 \ensuremath{\mu}l, (B) 20 \ensuremath{\mu}l and (C) 200 \ensuremath{\mu}l pipette tips to create a cross in the cell layer. We
concluded that 20 \ensuremath{\mu}l pipette tips are the most suitable for consistent
injury induction. Scale bar: 20 mm.}
\label{fig:supp_figure_04}
\end{figure}
\clearpage
\section*{Suppl. Note: Mathematical Framework for Latent-State Representation of Neuroimmune Responses}

\subsection*{S1. Latent-state representation}

To provide a quantitative interpretation of the experimentally observed
neuroimmune responses, we formulate a latent-state framework that
represents stimulus-specific phenotype patterns using a small number of
representative response programs. Rather than explicitly modeling the
intracellular signaling cascade, the proposed framework describes how an
external perturbation is transformed into experimentally measurable
cytokine, activation-marker, and morphological responses through an
intermediate latent-state representation. Let

\[
S\in\{\mathrm{LPS},\mathrm{poly}(\mathrm{I}\!:\!\mathrm{C}),\mathrm{scratch}\}
\]

denote the applied perturbation. The experimentally observed phenotype
vector is represented as

\[
\begin{aligned}\mathbf{Y}=(&Y_{\mathrm{TNF}\alpha},Y_{\mathrm{IL6}},Y_{\mathrm{IL1}\beta},Y_{\mathrm{IFN}\beta},\\ &Y_{\mathrm{IBA1}},Y_{\mathrm{CD68}},Y_{\mathrm{Amoeboid}},Y_{\mathrm{Border}}).\end{aligned}
\]

where the components correspond to cytokine secretion, activation-marker
expression, and morphology measurements.

We assume that the Instead of assuming that each perturbation directly
generates the observed phenotype, we introduce an intermediate latent
response program

\[
k\in\{\mathcal{P},\mathcal{V},\mathcal{W}\},
\]

where

\[
\begin{aligned}\mathcal{P}&:\text{Pro-inflammatory program},\\\mathcal{V}&:\text{Interferon-dominant program},\\\mathcal{W}&:\text{Wound-response program}.\end{aligned}
\]

The latent variable summarizes the dominant intracellular response
elicited by the applied stimulus before it is expressed as
experimentally measurable phenotypes. Consequently, the
stimulus-response relationship is described by the probabilistic
graphical model

\[
S\longrightarrow K\longrightarrow\mathbf{Y}.
\]

Conditioned on the applied stimulus, the activation probability of each
response program is described by a probabilty distribution

\[
\pi_{k\mid s}=P_{K\mid S}(k\mid s),\qquad k\in\{\mathcal{P},\mathcal{V},\mathcal{W}\}.
\]

Each latent response program is characterized by a representative
phenotype signature

\[
\boldsymbol{\mu}_k=(\mu_{k,1},\ldots,\mu_{k,8}),
\]

which summarizes the characteristic cytokine, activation-marker, and
morphology profile associated with program \ensuremath{k}.

The reconstructed phenotype under stimulus \ensuremath{s} is therefore expressed as

\[
\widehat{\mathbf{Y}}(s)=\sum_{k\in\{\mathcal{P},\mathcal{V},\mathcal{W}\}}P_{K\mid S}(k\mid s)\,\boldsymbol{\mu}_k.
\]

This representation assumes that the experimentally observed phenotype
is generated by a weighted combination of a small number of canonical
response programs. The latent-state formulation therefore provides a
compact mathematical description of the dominant stimulus-specific
phenotype organization observed in the experiments while substantially
reducing the dimensionality of the measured phenotype space. The
estimation of the response-program signatures and the corresponding
activation probabilities is described in Supplementary Note S2.

\subsection*{S2. Estimation of latent response programs}

The representative phenotype signatures were constructed from the
normalized phenotype responses measured under the three stimulation
paradigms. Rather than corresponding to discrete biological cell states,
the latent response programs provide representative computational
signatures that summarize the dominant stimulus-specific phenotype
organization observed experimentally. Three latent response programs
were selected because they correspond to the three dominant phenotype
patterns consistently observed across the experimental stimulation
paradigms.

For each phenotype feature, larger normalized values indicate stronger
relative responses. Accordingly, the representative phenotype signatures
were defined as

\[
\begin{aligned}\boldsymbol{\mu}_{\mathcal P}&=(1.00,0.85,0.80,0.10,0.95,0.40,0.90,0.50),\\\boldsymbol{\mu}_{\mathcal V}&=(0.10,0.10,0.20,1.00,0.20,0.30,0.30,0.20),\\\boldsymbol{\mu}_{\mathcal W}&=(0.10,0.10,0.70,0.10,0.65,0.50,0.80,1.00),\end{aligned}
\]

where each element represents the normalized relative response strength
of the corresponding phenotype feature.

For each stimulation condition, the activation probability of each
latent response program was estimated according to the dominant
phenotype pattern observed experimentally. The probabilities were
normalized such that each row sums to one and can therefore be
interpreted as the relative occupancy of the three latent response
programs under each stimulation condition. This yielded the
stimulus-to-program probability matrix

\[
P(K\mid S)=\begin{bmatrix}0.88&0.06&0.06\\0.08&0.84&0.08\\0.06&0.06&0.88\end{bmatrix}.
\]

where each row corresponds to one stimulation condition and each column
denotes the probability of activating the corresponding latent response
program.

The dominant diagonal structure of this matrix indicates that LPS
primarily activates the pro-inflammatory program, poly(I:C)
predominantly activates the interferon-dominant program, and scratch
injury mainly activates the wound-response program. The smaller
off-diagonal probabilities account for phenotype overlap among different
stimulation paradigms, reflecting shared biological responses such as
IL-1\ensuremath{\beta} production and activation-marker expression.

\subsection*{S3. Phenotype reconstruction}

Using the latent-state representation introduced in Supplementary Note
S1 and the representative response programs estimated in Supplementary
Note S2, the phenotype matrix was reconstructed for each stimulation
condition. The reconstructed phenotype therefore represents the expected
relative response of all measured phenotype features under the
corresponding perturbation.

The reconstructed phenotype matrix is shown in Fig. 8B. Although the
phenotype space is represented by only three latent response programs,
the model preserves the dominant stimulus-specific phenotype
organization observed experimentally. In particular, LPS stimulation
remains associated with a strong inflammatory phenotype characterized by
elevated TNF-\ensuremath{\alpha}, IL-6, IL-1\ensuremath{\beta}, and IBA1 responses. Poly(I:C) stimulation
is reconstructed as an interferon-dominant phenotype with prominent
IFN-\ensuremath{\beta} expression, whereas scratch stimulation is represented by a
wound-associated phenotype characterized by increased IL-1\ensuremath{\beta}, amoeboid
morphology, and scratch response.

To quantify the reconstruction error, the residual phenotype matrix was
computed as

\[
\mathbf{E}=\mathbf{Y}-\widehat{\mathbf{Y}},
\]

where \ensuremath{\mathbf{Y}} and \(\widehat{\mathbf Y}\) denote the experimentally observed and reconstructed
phenotype matrices, respectively.

The residual matrix is presented in Fig. 8C. Most residual values remain
close to zero, indicating close agreement between the reconstructed and
experimentally observed phenotype patterns. Larger residuals are mainly
observed for phenotype features shared among multiple response programs,
reflecting the partial overlap between inflammatory and wound-associated
signaling responses. Nevertheless, these deviations remain small
relative to the overall dynamic range of the normalized phenotype space.

Overall, the reconstruction results demonstrate that the experimentally
observed neuroimmune phenotype organization can be accurately
represented by a small number of latent response programs. These results
indicate that the proposed latent-state framework captures the dominant
organization of the experimentally observed phenotype space while
substantially reducing its dimensionality.

\subsection*{S4. Quantitative model validation}

To quantitatively evaluate the agreement between the reconstructed and
experimentally observed phenotype patterns, three complementary
performance metrics were employed: the coefficient of determination
(\(R^2\)), the Pearson correlation coefficient (\ensuremath{r}), and the
root-mean-square error (RMSE).

\Needspace{6\baselineskip}

The coefficient of determination was calculated as

\[
R^2=1-\frac{\sum_{i=1}^N(y_i-\widehat y_i)^2}{\sum_{i=1}^N(y_i-\overline y)^2},
\]

where \(y_i\) and \(\widehat y_i\) denote the observed and
reconstructed phenotype values, respectively, \(\overline y\) denotes the mean of the
observed phenotype values, and \ensuremath{N} is the total number of phenotype
measurements.

\Needspace{6\baselineskip}

The Pearson correlation coefficient was computed as

\[
r=\frac{\sum_{i=1}^N(y_i-\overline y)(\widehat y_i-\overline{\widehat y})}{\sqrt{\sum_{i=1}^N(y_i-\overline y)^2}\sqrt{\sum_{i=1}^N(\widehat y_i-\overline{\widehat y})^2}},
\]

where \(\overline{\widehat y}\) denotes the mean of the reconstructed phenotype values.

\Needspace{6\baselineskip}

The reconstruction error was further quantified using the
root-mean-square error,

\[
\mathrm{RMSE}=\sqrt{\frac{1}{N}\sum_{i=1}^N(y_i-\widehat y_i)^2}.
\]

The comparison between reconstructed and experimentally observed
phenotype values is presented in Fig. 9. The reconstructed phenotype
values exhibit a strong linear relationship with the corresponding
experimental observations, indicating that the latent-state framework
accurately captures the dominant organization of the phenotype space.
These quantitative metrics are consistent with the residual analysis
presented in Supplementary Note S3.

The proposed framework achieved

\[
R^2=0.962,\qquad r=0.996,\qquad\mathrm{RMSE}=0.066,
\]

demonstrating that the latent-state representation captures the dominant
structure of the experimentally observed neuroimmune phenotype space
while providing a compact low-dimensional description of
stimulus-specific cellular responses.

\subsection*{S5. Information-theoretic interpretation}

The latent-state framework introduced in this work naturally admits an
information-theoretic interpretation. The proposed framework interprets
the transformation from external perturbations to latent response
programs as an information-processing system. In this representation,
the applied perturbation acts as the system input, whereas the latent
response program represents the corresponding internal cellular
response,

\[
S\longrightarrow K.
\]

The stimulus-to-program relationship is characterized by the conditional
probability distribution

\[
P(K\mid S),
\]

which was estimated from the experimentally observed phenotype patterns
as described in Supplementary Note S2.

\Needspace{6\baselineskip}

Assuming equal prior probabilities for the three stimulation paradigms,
the mutual information between the applied perturbation and the latent
response program is defined as

\[
I(S;K)=\sum_{s\in\mathcal S}\sum_{k\in\mathcal K}P(s,k)\log_2\frac{P(s,k)}{P(s)P(k)}.
\]

The mutual information quantifies how much information about the applied
perturbation is preserved by the latent response programs. A larger
value indicates that different stimulation paradigms activate more
distinguishable latent cellular responses, whereas a smaller value
implies greater overlap among the corresponding phenotype patterns.

For the estimated stimulus-to-program probability matrix, the mutual
information is

\[
I(S;K)\approx0.887\ \text{bits}.
\]

Since the three stimulation paradigms are assumed to be equally likely,
the corresponding stimulus entropy is

\[
H(S)=\log_2 3\approx1.585\ \text{bits},
\]

yielding a normalized information efficiency of

\[
\frac{I(S;K)}{H(S)}\approx0.559.
\]

This result indicates that approximately 56\% of the available stimulus
information is preserved by only three latent response programs. The
remaining uncertainty reflects the fact that several experimentally
measured phenotype features, including IL-1\ensuremath{\beta} and activation-marker
expression, are shared among multiple response programs and therefore
cannot uniquely distinguish different stimulation paradigms.

The information-theoretic interpretation complements the latent-state
framework by providing a quantitative measure of stimulus specificity.
Rather than describing the underlying biochemical signaling mechanisms,
it characterizes how effectively different perturbations can be
distinguished through their corresponding latent response programs. This
perspective establishes a principled basis for comparing computational
models of neuroimmune responses in future studies.

\clearpage\section*{References}

\begingroup\small\setlength{\parskip}{6pt}

\noindent\hangindent=1.5em\hangafter=1

Abutbul, S., J. Shapiro, I. Szaingurten-Solodkin, et al. 2012. ``TGF-\ensuremath{\beta}
Signaling Through SMAD2/3 Induces the Quiescent Microglial Phenotype
Within the CNS Environment.'' \emph{Glia} 60: 1160--71.

\noindent\hangindent=1.5em\hangafter=1

Allan, S. M., and N. J. Rothwell. 2001. ``Cytokines and Acute
Neurodegeneration.'' \emph{Nature Reviews Neuroscience} 2: 734--44.

\noindent\hangindent=1.5em\hangafter=1

Butovsky, O., M. P. Jedrychowski, C. S. Moore, et al. 2014.
``Identification of a Unique TGF-\ensuremath{\beta}- Dependent Molecular and Functional
Signature in Microglia.'' \emph{Nature Neuroscience} 17: 131--43.

\noindent\hangindent=1.5em\hangafter=1

Campos, Ana B., Sara Duarte-Silva, António F. Ambrósio, Patrícia Maciel,
and Bruno Fernandes. 2021. ``MorphData: Automating the Data Extraction
Process of Morphological Features of Microglial Cells in ImageJ.''
\emph{bioRxiv}, 2021.08.05.455282.
\url{https://doi.org/10.1101/2021.08.05.455282}.

\noindent\hangindent=1.5em\hangafter=1

Ching, Travers, Daniel S Himmelstein, Brett K Beaulieu-Jones, et al.
2018. ``Opportunities and Obstacles for Deep Learning in Biology and
Medicine.'' \emph{Journal of the Royal Society Interface} 15 (141):
20170387.

\noindent\hangindent=1.5em\hangafter=1

Colonna, M., and O. Butovsky. 2017. ``Microglia Function in the Central
Nervous System During Health and Neurodegeneration.'' \emph{Annual
Review of Immunology} 35: 441--68.

\noindent\hangindent=1.5em\hangafter=1

Deczkowska, Aleksandra, Hadas Keren-Shaul, Assaf Weiner, Marco Colonna,
Michal Schwartz, and Ido Amit. 2018. ``Disease-Associated Microglia: A
Universal Immune Sensor of Neurodegeneration.'' \emph{Cell} 173 (5):
1073--81.

\noindent\hangindent=1.5em\hangafter=1

Erö, C., M. O. Gewaltig, D. Keller, and H. Markram. 2018. ``A Cell Atlas
for the Mouse Brain.'' \emph{Frontiers in Neuroinformatics} 12: 84.

\noindent\hangindent=1.5em\hangafter=1

Goshi, N., R. K. Morgan, P. J. Lein, and E. Seker. 2020. ``A Primary
Neural Cell Culture Model to Study Neuron, Astrocyte, and Microglia
Interactions in Neuroinflammation.'' \emph{Journal of Neuroinflammation}
17: 155.

\noindent\hangindent=1.5em\hangafter=1

He, Y., N. Taylor, X. Yao, and A. Bhattacharya. 2021. ``Mouse Primary
Microglia Respond Differently to LPS and Poly(i:c) in Vitro.''
\emph{Scientific Reports} 11: 10447.

\noindent\hangindent=1.5em\hangafter=1

Kaech, S., and G. Banker. 2006. ``Culturing Hippocampal Neurons.''
\emph{Nature Protocols} 1: 2406--15.

\noindent\hangindent=1.5em\hangafter=1

Karperien, Audrey L. 2007. \emph{FracLac for ImageJ}. \url{https://imagej.net/ij/plugins/fraclac/FLHelp/Introduction.htm}.

\noindent\hangindent=1.5em\hangafter=1

Lawson, L. J., V. H. Perry, P. Dri, and S. Gordon. 1990. ``Heterogeneity
in the Distribution and Morphology of Microglia in the Normal Adult
Mouse Brain.'' \emph{Neuroscience} 39: 151--70.

\noindent\hangindent=1.5em\hangafter=1

Levet, Florian, Johannes Tønnesen, U. Valentin Nägerl, and Jean-Baptiste
Sibarita. 2020. ``SpineJ: A Software Tool for Quantitative Analysis of
Nanoscale Spine Morphology.'' \emph{Methods} 174: 49--55.
\url{https://doi.org/10.1016/j.ymeth.2020.01.020}.

\noindent\hangindent=1.5em\hangafter=1

Liang, Chao-Cheng, Alice Y. Park, and Jun-Lin Guan. 2007. ``In Vitro
Scratch Assay: A Convenient and Inexpensive Method for Analysis of Cell
Migration in Vitro.'' \emph{Nature Protocols} 2 (2): 329--33.
\url{https://doi.org/10.1038/nprot.2007.30}.

\noindent\hangindent=1.5em\hangafter=1

Lopez-Castejon, G., and D. Brough. 2011. ``Understanding the Mechanism
of IL-1\ensuremath{\beta} Secretion.'' \emph{Cytokine \& Growth Factor Reviews} 22:
189--95.

\noindent\hangindent=1.5em\hangafter=1

Luecken, Malte D, and Fabian J Theis. 2019. ``Current Best Practices in
Single-Cell RNA-Seq Analysis: A Tutorial.'' \emph{Molecular Systems
Biology} 15 (6): MSB188746.

\noindent\hangindent=1.5em\hangafter=1

Mahley, R. W. 2016. ``Central Nervous System Lipoproteins: ApoE and
Regulation of Cholesterol Metabolism.'' \emph{Arteriosclerosis,
Thrombosis, and Vascular Biology} 36: 1305--15.

\noindent\hangindent=1.5em\hangafter=1

Masuda, Takahiro, Roman Sankowski, Ori Staszewski, and Marco Prinz.
2020. ``Microglia Heterogeneity in the Single-Cell Era.'' \emph{Cell
Reports} 30 (5): 1271--81.

\noindent\hangindent=1.5em\hangafter=1

Masuda, T., R. Sankowski, O. Staszewski, et al. 2019. ``Spatial and
Temporal Heterogeneity of Mouse and Human Microglia at Single-Cell
Resolution.'' \emph{Nature} 566: 388--92.

\noindent\hangindent=1.5em\hangafter=1

Michaelsen, Katrin, Masha Zagrebelsky, Jana Berndt-Huch, et al. 2010.
``Neurotrophin Receptors TrkB.T1 and p75NTR Cooperate in Modulating Both
Functional and Structural Plasticity in Mature Hippocampal Neurons.''
\emph{European Journal of Neuroscience} 32 (11): 1854--65.
\url{https://doi.org/10.1111/j.1460-9568.2010.07460.x}.

\noindent\hangindent=1.5em\hangafter=1

Paolicelli, Rosa C, Amanda Sierra, Beth Stevens, et al. 2022.
``Microglia States and Nomenclature: A Field at Its Crossroads.''
\emph{Neuron} 110 (21): 3458--83.

\noindent\hangindent=1.5em\hangafter=1

Saunders, A., E. Macosko, A. Wysoker, et al. 2018. ``A Single-Cell Atlas
of Cell Types, States, and Other Transcriptional Patterns from Nine
Regions of the Adult Mouse Brain.'' \emph{bioRxiv}, 299081.

\noindent\hangindent=1.5em\hangafter=1

Schindelin, J., I. Arganda-Carreras, E. Frise, et al. 2012. ``Fiji: An
Open-Source Platform for Biological-Image Analysis.'' \emph{Nature
Methods} 9: 676--82.

\noindent\hangindent=1.5em\hangafter=1

Schiweck, J., K. Murk, J. Ledderose, et al. 2021. ``Drebrin Controls
Scar Formation and Astrocyte Reactivity Upon Traumatic Brain Injury by
Regulating Membrane Trafficking.'' \emph{Nature Communications} 12:
1490.

\noindent\hangindent=1.5em\hangafter=1

Schmidt, Uwe, Martin Weigert, Coleman Broaddus, and Gene Myers. 2018.
``Cell Detection with Star-Convex Polygons.'' \emph{Medical Image
Computing and Computer Assisted Intervention -- MICCAI} \emph{2018},
Lecture notes in computer science, vol. 11071: 265--73.
\url{https://doi.org/10.1007/978-3-030-00934-2_30}.

\noindent\hangindent=1.5em\hangafter=1

Tejera, David, and Michael T. Heneka. 2016. ``Microglia in Alzheimer's
Disease: The Good, the Bad and the Ugly.'' \emph{Current}
\emph{Alzheimer} \emph{Research} 13 (4): 370--80.
\url{https://doi.org/10.2174/1567205013666151116120959}.

\noindent\hangindent=1.5em\hangafter=1

Valério-Gomes, B., D. M. Guimarães, D. Szczupak, and R. Lent. 2018.
``The Absolute Number of Oligodendrocytes in the Adult Mouse Brain.''
\emph{Frontiers in Neuroanatomy} 12: 90.

\noindent\hangindent=1.5em\hangafter=1

Wang, Y., K. J. Szretter, W. Vermi, et al. 2012. ``IL-34 Is a
Tissue-Restricted Ligand of CSF1R Required for the Development of
Langerhans Cells and Microglia.'' \emph{Nature Immunology} 13: 753--60.

\noindent\hangindent=1.5em\hangafter=1

Yao, Z., C. T. J. van Velthoven, M. Kunst, et al. 2023. ``A
High-Resolution Transcriptomic and Spatial Atlas of Cell Types in the
Whole Mouse Brain.'' \emph{Nature} 624: 317--32.

\endgroup

\end{document}